\documentclass[sn-mathphys]{sn-jnl}% Math and Physical Sciences Reference Style
\newtheorem{Remark}{Remark}

\newcommand{\argmin}{\mathop{\rm arg\min}}

\newcommand{\bbR}{\mathbb{R}}

\newcommand{\bbE}{{\mathbb{E}}}
\newcommand{\PP}{{\mathbb{P}}}

\newcommand{\Var}{{\rm Var}}

\newcommand{\Sigmab}{{{\Sigma}}}

\newcommand{\bX}{X}

\newcommand{\xnew}{x_{\rm new}}

\newcommand{\rmQ}{{\rm Q}}

\def\limsup{\mathop{\overline{\rm lim}}}

\newcommand{\setG}{G}
\newcommand{\QW}{{\rm Q}_{\Sigma}}
\newcommand{\QWh}{\widehat{{\rm Q}}_{\Sigma}} % with hat
\newcommand{\QA}{{\rm Q}_{A} }
\newcommand{\QAh}{\widehat{{\rm Q}}_{A} } % with hat

\newcommand{\VWh}{\widehat{{\rm V}}_{\Sigma}(\tau)} % with hat

\newcommand{\VAh}{\widehat{{\rm V}}_{A}(\tau)}

\newcommand{\betainit}{\widehat{\beta}}

\def\ep{\textsf{E}} % the symbol E for expectation used the sans serif letter
\def\Cov{\textsf{Cov}} % the symbol Cov for covariance used the sans serif letter
\def\Var{\textsf{Var}} % the symbol Var for covariance used the sans serif letter

\newcommand{\bI}{\mathbb{I}}
\newcommand{\beq}{\begin{equation}}
	\newcommand{\eeq}{\end{equation}}
\newcommand{\beas}{\begin{eqnarray*}}
	\newcommand{\eeas}{\end{eqnarray*}}
\newcommand{\bea}{\begin{eqnarray}}
	\newcommand{\eea}{\end{eqnarray}}

\def\th{\scalebox{0.8} {th}}
\def\cH{\mathcal{H}}
\def\one{\scriptscriptstyle (1)}
\def\two{\scriptscriptstyle (2)}
\def\ssd{\scriptscriptstyle (d)}
\def\td{\scalebox{0.4} {(d)}}

\jyear{2021}%

\theoremstyle{thmstyleone}%
\theoremstyle{thmstyletwo}%

\theoremstyle{thmstylethree}%

\begin{document}
	
\title[High-dimensional Regression]{Statistical Inference and Large-scale Multiple Testing for High-dimensional Regression Models}
	
	
\author[1]{\fnm{T. Tony} \sur{ Cai}}\email{tcai@wharton.upenn.edu}
	
\author[2]{\fnm{Zijian} \sur{ Guo}}\email{zijguo@stat.rutgers.edu}
	
\author[3]{\fnm{Yin} \sur{ Xia}}\email{xiayin@fudan.edu.cn}
	
\affil[1]{Department of Statistics and Data Science, University of Pennsylvania, Philadelphia, USA}
	
\affil[2]{Department of Statistics, Rutgers University, Piscataway, USA}
	
\affil[3]{Department of Statistics and Data Science, Fudan University, Shanghai, China}
	
	
\abstract{This paper presents a selective survey of recent developments in statistical inference and multiple testing for high-dimensional regression models, including linear and logistic regression. We examine the construction of confidence intervals and hypothesis tests for various low-dimensional objectives such as regression coefficients and linear and quadratic functionals. The key technique is to generate debiased and desparsified estimators for the targeted low-dimensional objectives and estimate their uncertainty. In addition to covering the motivations for and intuitions behind these statistical methods, we also discuss their optimality and adaptivity in the context of high-dimensional inference.  In addition, we review the recent development of statistical inference based on multiple regression models and the advancement of large-scale multiple testing for high-dimensional regression. The R package {\tt SIHR} has implemented some of the high-dimensional inference methods discussed in this paper.}

	
	
	
\maketitle

%\tableofcontents
%%%%%%%%%%%%%%%%%%%%%%%%%%%%
%%%%%%%%%%%%%%%%%%%%%%%%%%%%
\section{Introduction}\label{intro.sec}
%%%%%%%%%%%%%%%%%%%%%%%%%%%%
%%%%%%%%%%%%%%%%%%%%%%%%%%%%
%%%%%%%%%%%%%%%%%%%%%%%%%%%%%%%%%%%%%%%%%%%%%%%%%%%%%%%%%%%%
High-dimensional data analysis has become a vital part of scientific research in many fields. While the abundance of high-dimensional data offers numerous opportunities for statistical analysis, it also presents significant technical challenges, as the number of variables can be much larger than the number of observations. In high-dimensional settings, most of the classical inferential procedures, such as the maximum likelihood, are no longer valid. In recent years, there has been significant progress in developing new theories and methods for parameter estimation, hypothesis testing, confidence interval construction, and large-scale simultaneous inference in the context of high-dimensional data analysis. 

In this paper, we provide a selective survey of recent advances in statistical inference and multiple testing for high-dimensional regression models, commonly used in modern data analysis across various fields such as genetics, metabolomics, finance, health, and economics. 
Much progress has been made in estimation and prediction under the high-dimensional linear and generalized linear models (GLMs); see, for example, \cite{tibshirani1996regression,candes2007dantzig,zou2005regularization,zou2006adaptive,bickel2009simultaneous,buhlmann2011statistics,negahban2009unified,van2009conditions,huang2012estimation, fan2001variable,zhang2010nearly,belloni2011square,sun2012scaled,bunea2008honest,bach2010self,meier2008group,bellec2018slope,friedman2010regularization,efron2004least,greenshtein2004persistence}. Theoretical properties have been established in different settings, including the minimax estimation rate and the rate of convergence for the estimation and prediction errors of several penalized procedures. 

Uncertainty quantification is at the heart of many critical scientific applications and is more challenging than point estimation. For high-dimensional regression models, although the Lasso and other penalized estimators have been shown to achieve the optimal rates of convergence for estimation,
these estimators suffer from non-negligible bias that makes them unsuitable to be directly used for statistical inference, as noted in several studies \citep{van2014asymptotically,javanmard2014confidence,zhang2014confidence}. To overcome this issue, debiased inference methods have been developed in \citep{van2014asymptotically,javanmard2014confidence,zhang2014confidence} that correct the bias of penalized estimators and allow for statistical inference based on the debiased estimators. %This approach has been used to correct the bias of the Ridge estimator for statistical inference in \cite{buhlmann2013statistical}. 
The development of debiased inference methods has led to an increase in research on statistical inference for a wide range of low-dimensional objectives in different high-dimensional models; see, for example, \cite{guo2021inference,guo2019optimal,guo2021group,cai2020semisupervised,ma2022statistical,cai2017confidence,athey2018approximate,cai2021statistical,ning2017general,ren2015asymptotic,yu2018confidence,fang2020test,zhou2020estimation,zhao2014general,javanmard2020flexible,chen2019inference,zhu2018linear,guo2018testing,cai2018high,eftekhari2021inference,deshpande2018accurate,fang2017testing,neykov2018unified,dezeure2015high}.
In particular, \cite{cai2017confidence} studied the minimaxity and adaptivity of confidence intervals for general linear functionals of a high-dimensional regression vector and found significant differences between the cases of sparse and dense loading vectors. Another important inference tool, known as Neyman's orthogonalization or double machine learning, has been proposed in econometrics to enable inference for low-dimensional objectives with high-dimensional nuisance parameters; see, for example, \citep{belloni2014inference,chernozhukov2015valid,farrell2015robust,chernozhukov2018double,belloni2017program}. 

In the single regression model (one-sample) setting, we observe data $\{Y_k,X_{k,\cdot}\}_{1\leq k\leq n}$, where $Y_k\in \bbR$ and $X_{k,\cdot}\in \bbR^{p+1}$ denote the outcome and the high-dimensional covariates respectively, generated independently from the high-dimensional GLM, 
\setlength{\belowdisplayskip}{3pt} \setlength{\belowdisplayshortskip}{3pt}
\setlength{\abovedisplayskip}{3pt} \setlength{\abovedisplayshortskip}{3pt}
\begin{equation}
	Y_{k}=h(X_{k,\cdot}^{\intercal} \beta)+\epsilon_k, \quad \text{for}\quad 1\leq k\leq n
	\label{eq: GLM}
\end{equation}
with $\bbE(\epsilon_k\vert X_{k,\cdot})=0$ and the high-dimensional regression vector $\beta\in \bbR^{p+1}$. Throughout the paper, we use $\beta_1$ to denote the intercept and set $X_{k,1}=1$. We assume that the high-dimensional covariates $X_{k,-1} \in \bbR^{p}$ are centered and sub-gaussian and the matrix $\Sigma\equiv \bbE X_{k,\cdot} X_{k,\cdot}^{\intercal}\in \bbR^{(p+1)\times (p+1)}$ is well-conditioned. 
We focus on the linear and logistic regression models with the link function $h(z)=z$ and $h(z)=\exp(z)/[1+\exp(z)]$ respectively. The regression vector $\beta$ is assumed to be sparse, and its sparsity level is denoted by $\|\beta\|_0.$ The high-dimensional covariates $X_{k,\cdot}$ might come from a large number of measured covariates or the basis transformations of the baseline covariates. For the linear model, we further assume the error $\epsilon_k$ is sub-gaussian with homoscedastic regression error $\sigma_\epsilon^2=\bbE(\epsilon_k^2\vert X_{k,\cdot})$.

In addition to the one-sample setting, we examine the statistical inference methods for the two-sample high-dimensional regression models. 
For $d=1,2,$ we assume that the data $\{Y^{\ssd}_k,X^{\ssd}_{k,\cdot}\}_{1\leq k\leq n_d}$ are i.i.d. generated, following 
\begin{equation}
	\begin{aligned}
		Y^{\ssd}_k =h([X^{\ssd}_{k,\cdot}]^{\intercal}\beta^{\ssd})+\epsilon^{\ssd}_k, \quad \text{for}\quad 1\leq k\leq n_d, \end{aligned}
	\label{eq: multi-regression}
\end{equation}
where $\bbE(\epsilon^{\ssd}_k\vert X^{\ssd}_{k,\cdot})=0$ and $h(\cdot)$ is the pre-specified link function.

Based on the models above, this paper first addresses the challenges of making statistical inferences for low-dimensional objectives (e.g., linear and quadratic functionals) in high-dimensional regression, both in one- and two-sample settings. Specifically, the following quantities are of particular interest.
\begin{enumerate}
	\item {\bf Linear functional $\xnew^{\intercal}\beta$} with $\xnew\in\bbR^{p+1}$ in one-sample setting. The linear functional $\xnew^{\intercal}\beta$ includes as special cases the single regression coefficient $\beta_j$ \citep{van2014asymptotically, javanmard2014confidence, zhang2014confidence,cai2017confidence} when $\xnew$ is the $j^{\th}$ canonical unit vector and the conditional mean of the outcome under \eqref{eq: GLM} when $\xnew$ is a future observation's covariates. When $\xnew$ denotes the average of the covariates observations for a group, $\xnew^{\intercal}\beta$ is closely related to average treatment effect \cite{athey2018approximate}. In logistic regression, the linear functional $\xnew^{\intercal}\beta$ is closely related to the case probability \cite{guo2021inference}.
	
	\item {\bf Quadratic functionals $\beta_{G}^{\intercal}A\beta_{G}$ and $\beta_{G}^{\intercal}\Sigma_{G,G}\beta_{G}$} with $G\subset\{1,2,\cdots,p+1\}$ and $A\in \mathbb{R}^{\vert G \vert \times \vert G \vert}$ in one-sample setting. For a subset $G$, these quadratic functionals measure the total effect of variables in  $G.$ Statistical inference for quadratic functionals can be motivated from the group significance test, and the (local) genetic heritability estimation \cite{guo2019optimal,guo2021group}. The inference method can be generalized to handle heterogeneous effect tests, hierarchical testing, prediction loss evaluation, and confidence ball construction \cite{guo2021group,cai2020semisupervised}. 
	
	\item{\bf Difference between linear functionals $h(\xnew^{\intercal}\beta^{\two}) - h(\xnew^{\intercal}\beta^{\one})$} with $\xnew \in\bbR^{p+1}$ in two-sample setting. This difference measures the discrepancy between the conditional means, which is closely related to individual treatment selection for the new observation $\xnew \in \bbR^{p+1}$ \cite{cai2021optimal}.
	
	\item {\bf Inner products of regression vectors $[\beta^{\one}]^{\intercal}\beta^{\two}$ and $[\beta^{\one}]^{\intercal}A\beta^{\two}$} with the weighting matrix $A\in \bbR^{(p+1)\times (p+1)}$ in two-sample setting. The inner product of regression vectors or its weighted version measures the similarity between the two regression vectors. In genetic studies, such inner products can be used as the genetic relatedness measure when the covariates are genetic variants, and outcome variables are different phenotypes \cite{guo2019optimal,ma2022statistical}.
\end{enumerate}

We examine statistical inference procedures for linear and quadratic functionals from both methodological and theoretical perspectives. We also discuss the optimality results for the corresponding estimation and inference problems. A user-friendly R package \texttt{SIHR} \cite{rakshit2021sihr} has been developed to implement the statistical inference methods for the low-dimensional objectives mentioned above. This package provides a convenient way to apply the discussed statistical inference methods. 

Beyond the aforementioned inference for a single coordinate of the regression vector or other one-dimensional functionals, 
we also discuss the simultaneous inference of high-dimensional regression models. This includes using global methods with maximum-type statistics to test the entire regression coefficient vector \citep{dezeure2017high,zhang2017simultaneous,ma2021global}, as well as component-wise simultaneous inference methods that control the false discovery rate (FDR). Specifically, we examine the one-sample testing of high-dimensional linear regression coefficients \cite{liuluo2014}, the comparison of two high-dimensional linear regression models \cite{xia2018two}, and the joint testing of regression coefficients across multiple responses \cite{xia2018joint}. We also extend our discussion to logistic regression models. We discuss these large-scale multiple testing problems focusing on controlling the asymptotic FDR.

While error rate control is important for simultaneous inference, statistical power is also crucial. However, many existing testing methods for high-dimensional linear models do not consider auxiliary information, such as model sparsity and heteroscedasticity, that could improve statistical power. 
While there has been a significant amount of research on methods to enhance power in multiple testing in general \citep[among many others]{BenHoc97, storey2002direct, genovese2006false, Roeder2009, Ignatiadis2016,LeiFithian2018,Li2019,LAWS, fithian2022conditional}, recent efforts have also focused on simultaneous inference methods that incorporate auxiliary information to assist power improvement of high-dimensional regression analysis. For example, \cite{xia2018joint} achieved power gains by leveraging similarities across multivariate responses; \cite{xia2020gap} explored the sparsity information hidden in the data structures and improved the power through $p$-value weighting mechanisms; \cite{liu2021integrative} obtained power enhancement through integrating heterogeneous linear models. In the current paper, we primarily focus on power enhancement in a two-sample inference setting where the high-dimensional objects of interest are individually sparse. We will discuss methods for controlling FDR with and without power enhancement and the related theoretical aspects.

The rest of the paper is organized as follows. We finish this section by introducing the notation. Section \ref{sec: general idea}  discusses the debiased inference idea for the regression coefficients, and Section \ref{sec: functional} presents the debiased methods for linear and quadratic functionals in one- and two-sample settings. Section \ref{sec.multiple} focuses on simultaneous inference for high-dimensional regression models. We conclude the paper by discussing other related works in Section \ref{sec: discussion}. 

\noindent {\bf Notation.} For an event $E$, denote by $\bI\{E\}$ its indicator function. For an index set $J\subset\{1,2,\cdots,p\}$ and a vector $x\in \bbR^{p}$, $x_{J}$ is the sub-vector of $x$ with indices in $J$ and $x_{-J}$ is the sub-vector with indices in $J^{c}$. For a set $S$, $\vert S \vert$ denotes the cardinality of $S$. {For a vector $x\in \bbR^{p}$, the $\ell_q$ norm of $x$ is defined as $\|x\|_{q}=\left(\sum_{l=1}^{p}\vert x_l\vert ^q\right)^{\frac{1}{q}}$ for $q \geq 0$ with $\|x\|_0=\vert\{1\leq l\leq p: x_l\neq 0\}\vert$ and $\|x\|_{\infty}=\max_{1\leq l \leq p}\vert x_l\vert $.} Denote by $e_j$ the $j^{\th}$ canonical unit vector and ${\rm I}_{p} \in \bbR^{n \times p}$ the identity matrix. For a symmetric matrix $A$, denote by $\lambda_{\max}(A)$ and $\lambda_{\min}(A)$  its maximum and minimum eigenvalues, respectively.  For a matrix $A \in \bbR^{n \times p}$, $A_{\cdot,j}$ and $A_{i, \cdot}$ respectively denote the $j^{\th}$ column and $i^{\th}$ row of $A$, $A_{i,j}$ denotes the $(i,j)^{\th}$ entry of $A$, $A_{i,-j}$ denotes the $i^{\th}$ row of $A$ with its $j^{\th}$ entry removed, $A_{-i,j}$ denotes the $j^{\th}$ column of $A$ with its $i^{\th}$ entry removed, $A_{i,-\{j_1,j_2\}}$ denotes the $i^{\th}$ row of $A$ with its $j_1^{\th}$ and $j_2^{\th}$ entries both removed and $A_{-i,-j}\in \bbR^{(n-1)\times(p-1)}$ denotes the submatrix of $A$ with its $i^{\th}$ row and $j^{\th}$ column removed.
We use $c$ and $C$ to denote generic positive constants that may vary from place to place. For two positive sequences $a_n$ and $b_n$, $a_n \lesssim b_n$ means there exists a constant $C > 0$ such that $a_n \leq C b_n$ for all $n$;
$a_n \asymp b_n $ if $a_n \lesssim b_n$ and $b_n \lesssim a_n$, and $a_n \ll b_n$ if $\limsup_{n\rightarrow\infty} {a_n}/{b_n}=0$. 
Let $o_{\PP}\{a_n\}$ and $O_{\PP}\{a_n\}$ respectively represent the sequences that grow in a smaller and equal/smaller rate of the sequence $a_n$ with probability approaching $1$ as $n\rightarrow \infty$.

%%%%%%%%%%%%%%%%%%%%%%%%%%%%
%%%%%%%%%%%%%%%%%%%%%%%%%%%%
\section{Inference for Regression Coefficients}
\label{sec: general idea}
%%%%%%%%%%%%%%%%%%%%%%%%%%%%
%%%%%%%%%%%%%%%%%%%%%%%%%%%%

In this section, we begin with a discussion in Section \ref{sec: estimation} on several commonly used penalized estimators for the high-dimensional GLMs. We review in Section \ref{sec: debias linear} the debiased methods for linear models introduced in 
\cite{zhang2014confidence,van2014asymptotically,javanmard2014confidence} and discuss its extensions to high-dimensional logistic regression in Section \ref{sec: debias logistic}. We also present the optimality of the confidence interval construction in both linear and logistic high-dimensional regression models.

%%%%%%%%%%%%%%%%%%%%%%%%%%%%
%%%%%%%%%%%%%%%%%%%%%%%%%%%%
\subsection{Estimation in high-dimensional regression} 
\label{sec: estimation}
%%%%%%%%%%%%%%%%%%%%%%%%%%%%
%%%%%%%%%%%%%%%%%%%%%%%%%%%%
For the high-dimensional linear model \eqref{eq: GLM}, a commonly used estimator of the regression vector $\beta$ is the Lasso estimator \cite{tibshirani1996regression}, defined as
\begin{equation}
\small
	\betainit=\argmin_{\beta\in \bbR^{p+1}}\frac{\|Y-X\beta\|_2^2}{2n}+\lambda_0 \sum_{j=2}^{p+1} \frac{\|X_{\cdot, j}\|_2}{\sqrt{n}} \vert\beta_{j}\vert, 
	\label{eq: Lasso}
\end{equation}
with the tuning parameter $\lambda_0=A\sigma_{\epsilon}\sqrt{\log p/n}$ for some positive constant $A>2$. In the penalized regression \eqref{eq: Lasso}, we do not penalize the intercept $\beta_1$. The tuning parameter $\lambda_0$ is typically chosen by cross-validation, as implemented in the R package \texttt{glmnet} \cite{friedman2010regularization}.  With the Lasso estimator $\betainit,$ the variance $\sigma_\epsilon^2$ can be estimated by $\widehat{\sigma}_\epsilon^2=\tfrac{1}{n}\|Y-X\betainit\|^2_2.$ 

The tuning parameter $\lambda_0$ in \eqref{eq: Lasso} depends on the noise level $\sigma_{\epsilon}.$ Alternative estimators have been proposed such that the tuning parameter does not depend on the unknown noise \cite[e.g.]{sun2012scaled,belloni2011square}. Particularly, \cite{sun2012scaled} proposed the scaled Lasso estimator 
\begin{equation}
\small
	\{\betainit,\widehat{\sigma}_{\epsilon}\}=\arg\min_{\beta \in \bbR^{p+1},\sigma_\epsilon
		\in \bbR^{+}}\frac{\|y-X\beta\|_2^2}{2n\sigma_\epsilon}+\frac{\sigma_\epsilon}{2}+ \lambda_0 \sum_{j=2}^{p+1} \frac{\|X_{\cdot, j}\|_2}{\sqrt{n}} \vert\beta_j\vert 
	\label{eq: scaled Lasso}
\end{equation} 
with the tuning parameter $\lambda_0=A\sqrt{\log p/n}$ for some positive constant $A>2.$ Beyond the estimators mentioned above, a wide collection of estimators of high-dimensional regression vectors have been proposed \citep[e.g.,][]{zou2005regularization,zou2006adaptive,buhlmann2011statistics,negahban2009unified,huang2012estimation, fan2001variable,zhang2010nearly,belloni2011square}. 

For the high-dimensional logistic model \eqref{eq: GLM}, the penalized methods have also been well developed to estimate $\beta\in \bbR^{p+1}$
\citep[e.g.,][]{bunea2008honest,bach2010self,buhlmann2011statistics,meier2008group,negahban2009unified,huang2012estimation}. In this paper, we use the penalized log-likelihood estimator $\widehat{\beta}$ in \cite{buhlmann2011statistics}, defined as 
\begin{equation}
\small
	\widehat{\beta}=\argmin_{\beta}\frac{1}{n}\sum_{k=1}^{n}\{\log[1+\exp(X_{k,\cdot}^{\intercal}\beta)]-Y_k(X_{k,\cdot}^{\intercal}\beta)\}+\lambda_0 \sum_{j=2}^{p+1} \frac{\|X_{\cdot, j}\|_2}{\sqrt{n}} \vert\beta_j\vert,
	\label{eq: penalized MLE}
\end{equation}
with a positive tuning parameter $\lambda_0 \asymp \sqrt{\log p/n}$.

%%%%%%%%%%%%%%%%%%%%%%%%%%%%
%%%%%%%%%%%%%%%%%%%%%%%%%%%%
\subsection{Debiased or desparsified estimators in linear models}
\label{sec: debias linear}
%%%%%%%%%%%%%%%%%%%%%%%%%%%%
%%%%%%%%%%%%%%%%%%%%%%%%%%%%
The penalized estimators introduced in Section \ref{sec: estimation} have been shown to achieve the optimal convergence rate and satisfy desirable variable selection properties \cite{meinshausen2006high,bickel2009simultaneous,zhao2006model,wainwright2009sharp}. However, \citep{van2014asymptotically,javanmard2014confidence,zhang2014confidence} highlighted that the Lasso and other penalized estimators are not ready for statistical inference due to the non-negligible estimation bias. They further proposed correcting the penalized estimators' bias and then making inferences based on the bias-corrected estimators. 

In the following, we present the main idea of the bias correction method. To illustrate the main idea, we fix the parameter index $2\leq j\leq p+1$ and focus on the confidence interval construction for $\beta_j$ in the model \eqref{eq: GLM}. With $\widehat{\beta}$ denoting the Lasso estimator in \eqref{eq: Lasso}, the main idea of the method proposed in \cite{zhang2014confidence, javanmard2014confidence} is to estimate the error of the plug-in estimator $\betainit_j-\beta_j$. The approximation of the error $\betainit_j-\beta_j$ can be motivated by the following decomposition: for any vector $u\in  \bbR^{p+1}$,
\begin{equation}
\small
	{u^{\intercal}\frac{1}{n}\sum_{k=1}^{n} \bX_{k,\cdot}(Y_{k}-\bX_{k,\cdot}^{\intercal} \betainit)}-(\beta_j-\betainit_j)= {u^{\intercal}\frac{1}{n}\bX^{\intercal}\epsilon}+{\left(\widehat{\Sigma}u-e_j\right)^{\intercal}(\beta-\betainit)}, 
	\label{eq: bias estimation}
\end{equation}
with $\widehat{\Sigma}=\frac{1}{n}\sum_{k=1}^{n}X_{k,\cdot} X_{k,\cdot}^{\intercal}$. We explain how to construct the vector $u$ by balancing the two terms on the right-hand side of the decomposition in \eqref{eq: bias estimation}. 
The first term on the right-hand side of \eqref{eq: bias estimation} has the conditional variance $\sigma_\epsilon^2/n\cdot u^{\intercal} \widehat{\Sigma} u $ while the second term can be further upper-bounded by 
\begin{equation}
\small
	\left\vert\left(\widehat{\Sigma}u-e_j\right)^{\intercal}(\beta-\betainit)\right\vert\leq \|\widehat{\Sigma}u-e_j\|_{\infty}\|\beta-\betainit\|_1.
	\label{eq: remaining bias bound}
\end{equation}
An algorithmic method of constructing $u$ is to constrain the bias and minimize the variance. Particularly, \cite{zhang2014confidence,javanmard2014confidence} proposed the following construction of the projection direction, 
\begin{equation}
\small
	\widehat{u}=\;\argmin_{u\in \bbR^{p+1}} u^{\intercal} \widehat{\Sigma} u \quad \text{subject to}\quad 
	\|\widehat{\Sigma} u-e_j\|_{\infty}\leq \lambda
	\label{eq: projection linear}
\end{equation}
with $\lambda\asymp \sqrt{{\log p}/{n}}$ denoting a positive tuning parameter. The construction in \eqref{eq: projection linear} is designed to minimize the conditional variance $u^{\intercal} \widehat{\Sigma} u $ of the ``asymp normal" term and constrain $\|\widehat{\Sigma}u-e_j\|_{\infty}$, which further controls the ``remaining bias" term as in \eqref{eq: remaining bias bound}. 

With the projection direction $\widehat{u}$ in \eqref{eq: projection linear}, \cite{zhang2014confidence,javanmard2014confidence} introduced the debiased estimator, 
\begin{equation}
\small
	\widehat{\beta}^{\rm Deb}_j=\betainit_j+\widehat{u}^{\intercal}\frac{1}{n}\sum_{k=1}^{n} \bX_{k,\cdot}(Y_{k}-\bX_{k,\cdot}^{\intercal} \betainit), \quad \text{for}\quad 2\leq j\leq p+1.
	\label{eq: debiased estimator 0}
\end{equation}
Following from \eqref{eq: bias estimation}, we obtain the following error decomposition of $\widehat{\beta}^{\rm Deb}_j$,  
\begin{equation}
\small
	\widehat{\beta}^{\rm Deb}_j-\beta_j={\widehat{u}^{\intercal}\frac{1}{n}\sum_{k=1}^{n} \bX_{k,\cdot}\epsilon_k}+{(\widehat{\Sigma}\widehat{u}-e_j)^{\intercal}(\beta-\betainit)}.
	\label{eq: debiased decomp}
\end{equation}
The first term on the right-hand side of \eqref{eq: debiased decomp} is asymptotically normal with the asymptotic variance $(\sigma_\epsilon^2/n)\cdot \widehat{u}^{\intercal}\widehat{\Sigma}\widehat{u}$. Implied by \eqref{eq: remaining bias bound}, we show that the projection direction $\widehat{u}$ in \eqref{eq: projection linear} constrains the second term on the right-hand side of \eqref{eq: debiased decomp} by an upper bound $\lambda \|\beta-\betainit\|_1,$ with the tuning parameter $\lambda$ specified in \eqref{eq: projection linear}. This upper bound can be established to be of rate $\|\beta\|_0\log p/n$ and the rate of convergence of the estimating error $\widehat{\beta}^{\rm Deb}_j-\beta_j$ is $\frac{1}{\sqrt{n}}+\|\beta\|_0 \frac{\log p}{n},$
which is shown to be the minimax optimal rate of estimating the regression coefficient $\beta_j$ \cite{ren2015asymptotic,cai2021statistical}. More importantly, \cite{zhang2014confidence,javanmard2014confidence} have shown that the debiased estimator $\widehat{\beta}^{\rm Deb}_j$ in \eqref{eq: debiased estimator 0} is approximately unbiased and asymptotically normal when $\|\beta\|_0\ll \sqrt{n}/\log p$. Based on the asymptotic normality, \cite{zhang2014confidence,javanmard2014confidence} constructed the following confidence interval \begin{equation}
\small
	\mathrm{CI}=\left(\widehat{\beta}^{\rm Deb}_j-z_{\alpha / 2} \sqrt{\widehat{\mathrm{V}}}, \quad \widehat{\beta}^{\rm Deb}_j+z_{\alpha / 2} \sqrt{\widehat{\mathrm{V}}}\right) \quad \text{with}\quad \widehat{\mathrm{V}}=\frac{\widehat{\sigma}_{\epsilon}^{2}}{n} \widehat{u}^{\intercal} \widehat{\Sigma} \widehat{u},
	\label{eq: CI linear coef}
\end{equation}
where $z_{\alpha/2}$ is the upper $\alpha/2$ quantile for the standard normal distribution.

\begin{Remark}\rm In the low-dimensional setting, we may set $\lambda=0$ as in \eqref{eq: projection linear} and obtain $\widehat{u}=\widehat{\Sigma}^{-1}e_j.$ This choice of $\widehat{u}$ reduces the debiased estimator in \eqref{eq: debiased estimator 0} to the OLS estimator.  
	The debiased estimator is also referred to as the ``desparsified" estimator \cite{zhang2014confidence,van2014asymptotically} since $\widehat{\beta}^{\rm Deb}_j$ is generally not zero even if the true $\beta_j$ is zero. Hence, even if $\beta$ is a sparse vector, the vector $\widehat{\beta}^{\rm Deb}=(\widehat{\beta}^{\rm Deb}_1,\widehat{\beta}^{\rm Deb}_2,\cdots, \widehat{\beta}^{\rm Deb}_{p+1})^{\intercal}$ is dense. Consequently, the vector $\widehat{\beta}^{\rm Deb}$ does not estimate $\beta$ well in general even though $\widehat{\beta}^{\rm Deb}_j$ is an optimal estimator of $\beta_j$ for every $2\leq j\leq p+1.$
\end{Remark}

%%%%%%%%%%%%%%%%%%%%%%%%%%%%%%%%%%%%%%%%%%%%%%%%%%%%%%%%%%%%%%%%%%%%%%%%%%%%%%%%%%%%%%%%%%%%%%%%%%%%%%%%%%%%%%%%
\subsubsection{Optimality of statistical inference} 
\begin{figure}[H]
	\centering
	\includegraphics[scale=0.4]{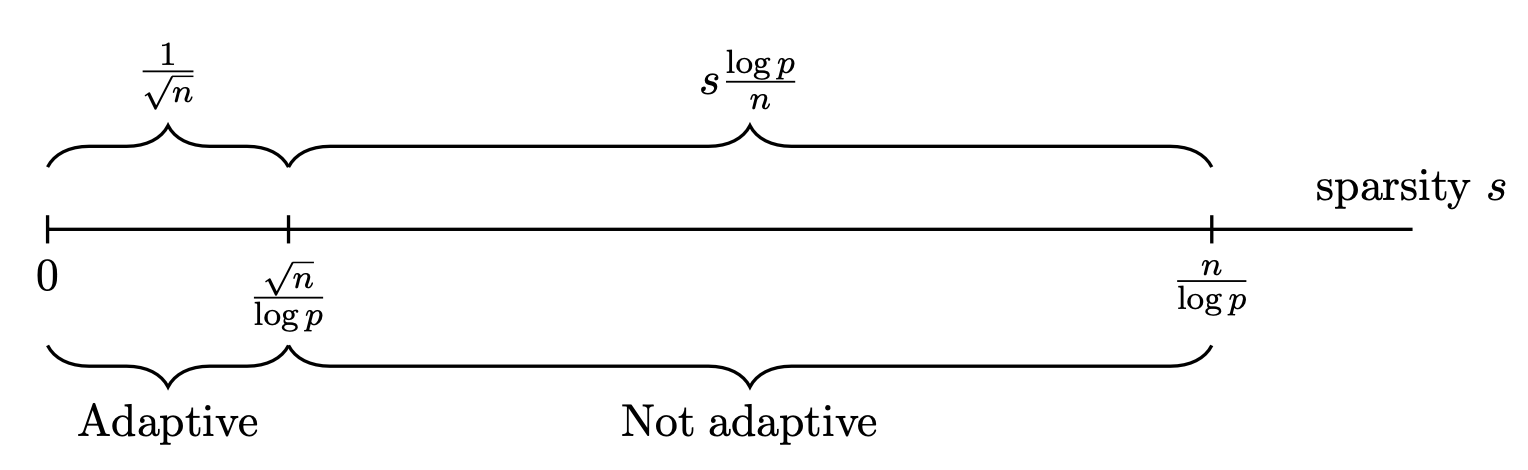}
	\caption{An illustration of the minimax optimality and adaptivity of the confidence intervals concerning the sparsity $s$ of $\beta$ for the unknown design setting. {On the top of the figure, we report the minimax expected lengths of the confidence intervals. At the bottom of the figure, the possibility of being adaptive to the sparsity $s$ is presented.} } 
	\label{fig: optimal}
\end{figure}

%%%%%%%%%%%%%%%%%%%%%%%%%%%%%%%%%%%%%%%%%%%%%%%%%%%%%%%%%%%%%%%%%%%%%%%%%%%%%%%%%%%%
In high-dimensional linear model, the paper \cite{cai2021statistical} established the minimax expected length of the confidence interval over the parameter space 
\begin{equation}
\small
	\Theta(s)=\left\{\theta=(\beta,\Sigma,\sigma_\epsilon):
	\|\beta\|_0\leq s, c_0\leq \lambda_{\min} (\Sigma)\leq \lambda_{\max}(\Sigma) \leq C_0, 0<\sigma_\epsilon\leq C_1\right\},
	\label{eq: parameter space}
\end{equation}
where $C_0\geq c_0>0$ and $C_1>0$ are positive constants. The space $\Theta(s)$ contains all regression vectors of less than $s$ non-zero elements. As established in \cite{cai2021statistical}, the minimax expected length of confidence intervals over $\Theta(s)$ for the regime $s \lesssim n/\log p$ is 
$\frac{1}{\sqrt{n}}+s\frac{\log p}{n}.$ When $s \lesssim \frac{\sqrt{n}}{\log p}$, the optimal length $1/\sqrt{n}$ can be achieved by the confidence interval in \eqref{eq: CI linear coef}. Over the regime $\frac{\sqrt{n}}{\log p}\ll s \lesssim n/\log p$, \cite{cai2021statistical} proposed a confidence interval attaining the optimal rate $s \log p/n$, where the construction requires the prior knowledge of the sparsity level $s.$ We illustrate the minimax expected length in Figure \ref{fig: optimal}. 
%%%%%%%%%%%%%%%%%%%%%%%%%%%%%%%
More importantly, \cite{cai2017confidence} studied the possibility of constructing a rate-optimal adaptive confidence interval. Here, an adaptive confidence interval means that, even without knowing about the sparsity level $s$, the length of the constructed confidence interval is automatically adapting to $s$. Since the sparsity information of the regression vector $\beta$ is generally unknown, it is desirable to construct an adaptive confidence interval. The work \cite{cai2017confidence} established the following adaptivity results: if the design covariance matrix $\Sigma$ is unknown, it is possible to construct an adaptive confidence interval only for the ultra-sparse regime $s \lesssim \sqrt{n}/\log p$. That is, if $\sqrt{n}/\log p\ll s \lesssim {n}/\log p$, it is impossible to construct a rate-optimal confidence interval that is adaptive to the sparsity level. This phase transition about the possibility of constructing adaptive confidence intervals is presented in Figure \ref{fig: optimal}.

The information of $\Sigma$ is critical for constructing optimal confidence intervals. If $\Sigma$ is known, the minimax expected length is $1/\sqrt{n}$ over the entire sparse regime $s \lesssim n/\log p$; see \cite{javanmard2018debiasing,cai2017confidence} for details. This contrasts sharply with the optimality results in Figure \ref{fig: optimal}.

\iffalse
\fi

%%%%%%%%%%%%%%%%%%%%%%%%%%%%%%%%%%%%%%%%%%%%%%%%%%%%%%%%%%%%%%%%%%%%%%%%%%%%%%%%%%%%%%%%%%%%%%%%%%%%%%%%%%
\subsubsection{Another viewpoint: debiasing with decorrelation}\label{debias2.sec}
%%%%%%%%%%%%%%%%%%%%%%%%%%%%%%%%%%%%%%%%%%%%%%%%%%%%%%%%%%%%%%%%%%%%%%%%%%%%%%%%%%%%%%%%%%%%%%%%%%%%%%%%%%

In this subsection, we detour to introduce a slightly different view of debiased Lasso estimator proposed in \cite{zhang2014confidence,van2014asymptotically}, which is of the following form 
\begin{equation}
\small
	\widetilde{\beta}^{\rm Deb}_j=\frac{Z_{\cdot,j}^{\intercal}(Y-X_{\cdot,-j}\betainit_{-j})}{Z_{\cdot,j}^{\intercal}X_{\cdot,j}},
	\label{eq: decorrelation}
\end{equation}
where $Z_{\cdot,j}\in \bbR^{n}$ is a decorrelation vector to be specified. \cite{zhang2014confidence,van2014asymptotically} proposed to construct the vector $Z_{\cdot,j}\in \bbR^{n}$ as the residual $Z_{\cdot,j}=X_{\cdot,j}-X_{\cdot,-j}\widehat{\gamma}$, with the Lasso estimator 
$\widehat{\gamma}= \frac{1}{2n}\|X_{\cdot,j}-X_{\cdot,-j}\gamma\|_2^2+\lambda_{\gamma} \sum_{l\neq j}\frac{\|X_{\cdot,l}\|_2}{\sqrt{n}} \vert\gamma_l\vert,$ where $\lambda_{\gamma}>0$ is a positive tuning parameter. The KKT condition ensures that the residual $Z_{\cdot,j}=X_{\cdot,j}-X_{\cdot,-j}\widehat{\gamma}$ is nearly orthogonal to all columns of $X_{\cdot,-j}$, via $${\small\frac{1}{n}\|Z_{\cdot,j}^{\intercal}X_{\cdot,-j}\|_{\infty}\leq \lambda_{\gamma} \cdot \max_{l\neq j} \frac{\|X_{\cdot,l}\|_2}{\sqrt{n}}.}$$  To see the effectiveness of the estimator in \eqref{eq: decorrelation}, we examine its estimation error 
\begin{equation}
\small
	\widetilde{\beta}^{\rm Deb}_j-\beta_j=\frac{Z_{\cdot,j}^{\intercal}\epsilon}{Z_{\cdot,j}^{\intercal}X_{\cdot,j}}+\frac{Z_{\cdot,j}^{\intercal}X_{\cdot,-j}(\beta_{-j}-\betainit_{-j})}{Z_{\cdot,j}^{\intercal}X_{\cdot,j}}.
	\label{eq: decorrelation alter}
\end{equation}
In the above expression, the first term on the right-hand side can be shown to be asymptotically normal under regularity conditions, while the second term is constrained as 
\begin{align*}
{\small
	\left\vert\frac{Z_{\cdot,j}^{\intercal}X_{\cdot,-j}(\beta_{-j}-\betainit_{-j})}{Z_{\cdot,j}^{\intercal}X_{\cdot,j}}\right\vert \leq \frac{1}{n}\|Z_{\cdot,j}^{\intercal}X_{\cdot,-j}\|_{\infty}\cdot \frac{\|\beta_{-j}-\betainit_{-j}\|_1}{\frac{1}{n}\vert Z_{\cdot,j}^{\intercal}X_{\cdot,j}\vert}\leq C \lambda_{\gamma} \|\widehat{\beta}-\beta\|_1,}
\end{align*}	
where the last inequality holds with a high probability for some positive constant $C>0.$ With the above argument, \cite{zhang2014confidence,van2014asymptotically} have shown that the first term in the decomposition \eqref{eq: decorrelation alter} is the dominating term for the ultra-sparse regime $\|\beta\|_0\ll \sqrt{n}/\log p$.
This leads to the asymptotic normality of the estimator $\widetilde{\beta}^{\rm Deb}_j$. \cite{zhang2014confidence,van2014asymptotically} further constructed the confidence interval based on asymptotic normality. In the current paper, we focus on methods generalizing the debiased estimator in \eqref{eq: debiased estimator 0}, instead of the decorrelation form in \eqref{eq: decorrelation}. However, the decorrelation idea has been extended to handle other statistical inference problems, including high-dimensional generalized linear model \cite{ning2017general,van2014asymptotically}, gaussian graphical model \cite{ning2017general}, high-dimensional confounding model \cite{guo2020doubly,sun2022decorrelating}, and high-dimensional additive model \cite{guo2019local}. 
 
%\begin{equation*}
%\small
%	\mathrm{CI}=\left(\widetilde{\beta}^{\rm Deb}_j-z_{\alpha / 2} \sqrt{\widetilde{\mathrm{V}}}, \quad \widetilde{\beta}^{\rm Deb}_j+z_{\alpha / 2} \sqrt{\widetilde{\mathrm{V}}}\right) \quad \text{with}\quad \widetilde{\rm V}=\frac{\widehat{\sigma}_{\epsilon}^{2}\|Z_{\cdot,j}\|_2^2}{(Z_{\cdot,j}^{\intercal}X_{\cdot,j})^2}.
%\end{equation*}
%The debiased estimator $\widetilde{\beta}^{\rm Deb}_j$ has a similar theoretical property as that of $\widehat{\beta}^{\rm Deb}_j$ in \eqref{eq: debiased estimator 0}. Particularly, if we set $Z_{\cdot,j}=X\widehat{u}$ in \eqref{eq: decorrelation}, then $\widetilde{\beta}^{\rm Deb}_j$ is the same as $\widehat{\beta}^{\rm Deb}_j.$

%%%%%%%%%%%%%%%%%%%%%%%%%%%%
%%%%%%%%%%%%%%%%%%%%%%%%%%%%
\subsection{Debiasing in binary outcome models}
\label{sec: debias logistic}
%%%%%%%%%%%%%%%%%%%%%%%%%%%%
%%%%%%%%%%%%%%%%%%%%%%%%%%%%
We generalize the debiased estimator in \eqref{eq: debiased estimator 0} to high-dimensional GLM with a binary outcome. Similarly to the Lasso estimator, the penalized logistic regression estimator $\widehat{\beta}$ in \eqref{eq: penalized MLE} suffers from the bias due to the $\ell_1$ penalty. The bias-corrected estimator is proposed as the following generic form, 
\begin{equation}
\small
	\widehat{\beta}^{\rm Deb}_j=\widehat{\beta}_j+\widehat{u}^{\intercal}\frac{1}{n}\sum_{k=1}^{n}W_kX_{k,\cdot}(Y_k-h(X_{k,\cdot}^{\intercal}\widehat{\beta})),
	\label{eq: generic form}
\end{equation}
where $W_k\in \bbR$, for $1\leq k\leq n$, are the weights to be specified and $\widehat{u}\in \bbR^{p+1}$ is the projection direction to be specified. We apply the Taylor expansion of the $h$ function and obtain
\begin{equation}
	\small
	\begin{aligned}
		Y_k - h(X_{k,\cdot}^{\intercal}\widehat{\beta}) = h(X_{k,\cdot}^{\intercal}\beta) - h(X_{k,\cdot}^{\intercal}\widehat{\beta}) + \epsilon_k = h^{\prime}(X_{k,\cdot}^{\intercal}\widehat{\beta})X_{k,\cdot}^{\intercal}(\beta - \widehat{\beta}) + R_k + \epsilon_k
	\end{aligned}
	\label{eq: taylor}
\end{equation}
with the approximation error {\small $R_k = \int_{0}^{1} (1-t)h''(X_{k,\cdot}^{\intercal}\widehat{\beta}+tX^{\intercal}_{k\cdot}(\beta-\widehat{\beta})) dt \cdot (X^{\intercal}_{k\cdot}(\widehat{\beta}-\beta))^2$}.
We plug in the Taylor expansion \eqref{eq: taylor} into the weighted bias-correction estimator in \eqref{eq: generic form}, leading to the error decomposition of $\widehat{\beta}^{\rm Deb}_j-\beta_j$ as 
\begin{equation}
\small
	\begin{aligned}
	&\underbrace{\frac{1}{n}\sum_{k=1}^{n}\widehat{u}^{\intercal}X_{k,\cdot}W_k\epsilon_k}_{\text{asymp normal}} + \underbrace{\frac{1}{n}\sum_{k=1}^{n}\left(W_k h^{\prime}(X_{k,\cdot}^{\intercal}\widehat{\beta})X_{k,\cdot}X_{k,\cdot}^{\intercal}\widehat{u}-e_j\right)^{\intercal}(\beta - \widehat{\beta})}_{\text{remaining bias}}\\
	&+\underbrace{\frac{1}{n}\sum_{k=1}^{n}\widehat{u}^{\intercal}X_{k,\cdot}W_k R_k}_{\text{nonlinearity bias}}.
	\end{aligned}
	\label{eq: key decomposition logistic}
\end{equation}

In the following, we describe two ways of specifying the weights $W_k$. \begin{enumerate}
	\item {\bf Linearization weighting.} \cite{guo2021inference} proposed to construct the weight 
	$W_k=1/h'(X_{k,\cdot}^{\intercal}\widehat{\beta})$. Then the ``remaining bias" term in \eqref{eq: key decomposition logistic} is reduced to $\frac{1}{n}\sum_{k=1}^{n}\left(X_{k,\cdot}X_{k,\cdot}^{\intercal}\widehat{u}-e_j\right)^{\intercal}(\beta - \widehat{\beta}),$ which is the same as the corresponding term in the linear regression. This enables us to directly adopt the projection direction $\widehat{u}$ constructed in \eqref{eq: projection linear}. This connection reveals the advantage of the weight $W_k=1/h'(X_{k,\cdot}^{\intercal}\widehat{\beta})$, that is, the bias-correction developed under the linear regression model can be directly extended to the logistic regression. 
	\item {\bf Link-specific weighting.} For a general link function $h(\cdot)$, \cite{cai2021statistical} constructed the weight $W_k=h'(X_{k,\cdot}^{\intercal}\widehat{\beta})/[h(X_{k,\cdot}^{\intercal}\widehat{\beta})(1-h(X_{k,\cdot}^{\intercal}\widehat{\beta}))]$. If $h$ is the logistic link, we have $h'(\cdot)=h(\cdot)(1-h(\cdot))$ and obtain the constant weight $W_k=1$. Such a link-specific weighting can be generalized to other binary outcome models (e.g., the probit model) with various link functions $h(\cdot)$; see details in \cite{cai2021statistical}.
\end{enumerate} 

After specifying the weights, the projection direction can be constructed as 
\begin{equation}
\small
	\begin{aligned}
		\widehat{u}&=\;\argmin_{u\in \bbR^{p+1}} u^{\intercal} \left(\frac{1}{n}\sum_{k=1}^{n} W_k h'(X_{k,\cdot}^{\intercal}\widehat{\beta})X_{k,\cdot}X_{k,\cdot}^{\intercal}\right) u \\
		\text{subject to}&\quad \left \|\left(\frac{1}{n}\sum_{k=1}^{n} W_k h'(X_{k,\cdot}^{\intercal}\widehat{\beta})X_{k,\cdot}X_{k,\cdot}^{\intercal}\right) u-e_j\right\|_{\infty}\leq \lambda, \quad \|Xu\|_{\infty}\leq \tau
	\end{aligned}
	\label{eq: projection logistic}
\end{equation}
%\Zijian{This is correct for the first weight, where we set $W_k=1/h'(X_{k,\cdot}^{\intercal}\widehat{\beta})$ and hence the whole optimization problem is reduced to being the same as the linear regression.}
with the positive tuning parameters $\lambda\asymp \sqrt{{\log p}/{n}}$ and $\tau \asymp \sqrt{\log n}$. The construction in \eqref{eq: projection logistic} can be motivated from a similar view as \eqref{eq: projection linear}. The constraint $\left \|\left(\frac{1}{n}\sum_{k=1}^{n} W_k h'(X_{k,\cdot}^{\intercal}\widehat{\beta})X_{k,\cdot}X_{k,\cdot}^{\intercal}\right) u-e_j\right\|_{\infty}\leq \lambda$ is imposed to constrain the ``remaining bias" term in \eqref{eq: key decomposition logistic} and $\|Xu\|_{\infty}\leq \tau$ is imposed to constrain the ``nonlinearity bias" in \eqref{eq: key decomposition logistic}. {For the linearization weighting, the conditional variance of $\frac{1}{n}\sum_{k=1}^{n}{u}^{\intercal}X_{k,\cdot}W_k\epsilon_k$ is $u^{\intercal} \left(\frac{1}{n}\sum_{k=1}^{n} W^2_k h'(X_{k,\cdot}^{\intercal}\widehat{\beta})X_{k,\cdot}X_{k,\cdot}^{\intercal}\right) u$, which is of the same order as the objective function in \eqref{eq: projection logistic} for a bounded $W_k.$ So, instead of minimizing the exact variance, we minimize a scaled conditional variance in \eqref{eq: projection logistic}, which has the advantage of leading to almost the same optimization as in \eqref{eq: projection linear} for the linear model. } 

Theoretical properties of the debiased estimators \eqref{eq: generic form} have been established for the logistic outcome model. With the weights $W_k=1/h'(X_{k,\cdot}^{\intercal}\widehat{\beta})$ for the linearization weighting, \cite{guo2021inference} established the asymptotic normality of $\widehat{\beta}^{\rm Deb}_j$ in \eqref{eq: generic form}. \cite{cai2021statistical} established a similar theoretical property for $\widehat{\beta}^{\rm Deb}_j$ in \eqref{eq: generic form} with the weights $W_k=1$ for the link-specific weighting. The asymptotic normality results in both works require the ultra-sparse condition $\|\beta\|_0\ll \sqrt{n}/[\log p\log n]$. %\Yin{I am wondering, for the following two weighting methods, is there any advantages of one method over the other, such as smaller variance, etc? Shall we discuss a bit?}
The use of the weights $W_k=1$ in \cite{cai2021statistical} leads to a smaller standard error than using the weights $W_k=1/h'(X_{k,\cdot}^{\intercal}\widehat{\beta})$ proposed in \cite{guo2021inference}.  The theoretical justification in \cite{cai2021statistical} requires a sample splitting such that the initial estimator $\widehat{\beta}$ is constructed from an independent sample. Such sample splitting is not required in the analysis of \cite{guo2021inference}, which is part of the benefit of linearization weighting. 

Based on the asymptotic normality, we construct the confidence interval 
\begin{equation}
\small
	\mathrm{CI}=\left(\widehat{\beta}^{\rm Deb}_j-z_{\alpha / 2} \sqrt{\widehat{\mathrm{V}}}, \quad \widehat{\beta}^{\rm Deb}_j+z_{\alpha / 2} \sqrt{\widehat{\mathrm{V}}}\right) \quad \text{with}\quad \widehat{\mathrm{V}}=\frac{1}{n} \widehat{u}^{\intercal} \widehat{\Sigma}^{G} \widehat{u},
	\label{eq: CI logistic coef}
\end{equation}
where $\widehat{\Sigma}^{G}=\frac{1}{n}\sum_{k=1}^{n} W_k^2h(X_{k,\cdot}^{\intercal}\widehat{\beta})(1-h(X_{k,\cdot}^{\intercal}\widehat{\beta}))X_{k,\cdot} X_{k,\cdot}^{\intercal}.$ %\Yin{Why do we need a superscript $G$ here? Is it a typo?} \Zijian{Not a typo since this is a general covariance matrix with the weight there.}

The optimality of confidence interval construction in high-dimensional logistic regression was studied in \cite{cai2021statistical}. The minimax expected length and the possible regime of constructing adaptive confidence intervals are similar to those in Figure \ref{fig: optimal}, up to a polynomial order of $\log n$; see the results in \cite{cai2021statistical}.

%\Zijian{Stop here.}
%%%%%%%%%%%%%%%%%%%%%%%%%%%%
%%%%%%%%%%%%%%%%%%%%%%%%%%%%
\section{Linear and Quadratic Functionals Inference}
\label{sec: functional}
%%%%%%%%%%%%%%%%%%%%%%%%%%%%
%%%%%%%%%%%%%%%%%%%%%%%%%%%%

We consider in this section statistical inference for linear and quadratic transformations of the regression vectors under high-dimensional linear and logistic regressions. We investigate both one- and two-sample regression models. In Section \ref{sec: package}, we discuss the R package \texttt{SIHR} \cite{rakshit2021sihr} that implements these methods. 

%%%%%%%%%%%%%%%%%%%%%%%%%%%%
%%%%%%%%%%%%%%%%%%%%%%%%%%%%
\subsection{Linear functionals for linear regression}
\label{sec: LF linear}
%%%%%%%%%%%%%%%%%%%%%%%%%%%%
%%%%%%%%%%%%%%%%%%%%%%%%%%%%
For a given vector $\xnew\in \bbR^{p+1}$, we present the construction of the point estimator and confidence interval for $x_{\text{new}}^{\intercal}\beta$ under the high-dimensional linear model \eqref{eq: GLM}. Similar to inference for $\beta_j$, the plug-in estimator $\xnew^{\intercal}\betainit$ suffers from the estimation bias by directly plugging in the Lasso estimator $\betainit$ in \eqref{eq: Lasso}. The work \cite{cai2021optimal} proposed the following bias-corrected estimator,
\begin{equation}
\small
	\widehat{x_{\rm new}^{\intercal}\beta}=x_{\rm new}^{\intercal}\betainit+\widehat{u}^{\intercal} \frac{1}{n}\sum_{k=1}^{n} \bX_{k,\cdot}(Y_{k}-\bX_{k,\cdot}^{\intercal} \betainit), 
	\label{eq: LF linear}
\end{equation}
with the projection direction $\widehat{u}$ defined as 
\begin{align}
\small
	\widehat{u}=\;\argmin_{u\in \bbR^{p+1}} u^{\intercal} \widehat{\Sigma} u \quad \text{subject to}\;
	& \left \|\widehat{\Sigma} u-x_{\rm new}\right\|_{\infty}\leq \|x_{\rm new}\|_2 \lambda \label{eq: constraint 1}\\
	&\; \left\vert x_{\rm new}^{\intercal} \widehat{\Sigma} u-\|x_{\rm new}\|_2^2 \right\vert\leq \|x_{\rm new}\|_2^2\lambda, \label{eq: constraint 2}
\end{align}
where $\widehat{\Sigmab}=\frac{1}{n}\sum_{k=1}^{n}X_{k,\cdot} X_{k,\cdot}^{\intercal}$ and $\lambda\asymp \sqrt{{\log p}/{n}}$ is a positive tuning parameter.

The debiased estimator in \eqref{eq: LF linear} satisfies the following error decomposition, 
\begin{equation}
\small
	\widehat{x_{\rm new}^{\intercal}\beta}-x_{\rm new}^{\intercal}\beta=\underbrace{\widehat{u}^{\intercal}\frac{1}{n}\bX^{\intercal}\epsilon}_{\text{asymp normal}}+\underbrace{\left(\widehat{\Sigma} \widehat{u}-\xnew\right)^{\intercal}(\beta-\betainit)}_{\text{remaining bias}}.
	\label{eq: decomp LF}
\end{equation}

The construction in \eqref{eq: constraint 1}, without the additional constraint \eqref{eq: constraint 2}, can be viewed as a direct generalization of \eqref{eq: projection linear} by replacing $e_j$ with the general loading $\xnew$. Specifically, \eqref{eq: constraint 1} minimizes the conditional variance of the ``asymp normal" term in \eqref{eq: decomp LF} and provides a control of the ``remaining bias" term by the inequality $\vert(\widehat{\Sigma} u-\xnew)^{\intercal}(\beta-\betainit)\vert\leq \|\widehat{\Sigma} u-x_{\rm new}\|_{\infty}\|\beta-\betainit\|_1.$ Such a construction of the projection direction for linear functional has been proposed in \cite{cai2017confidence,athey2018approximate}. However, such a direct generalization is not universally effective for all loadings. As established in Proposition 2 in \cite{cai2021optimal}, the projection direction, constructed without the additional constraint \eqref{eq: constraint 2}, does not correct the bias for a wide class of dense loading vectors.  

We shall emphasize that the new constraint in \eqref{eq: constraint 2} is crucial to ensuring the asymptotic normality of $\widehat{x_{\rm new}^{\intercal}\beta}-x_{\rm new}^{\intercal}\beta$ for any loading vector $\xnew.$ This constraint is imposed such that the variance of the ``asymp normal" term in \eqref{eq: decomp LF} always dominates the ``remaining bias" term in \eqref{eq: decomp LF}. With this additional constraint, the projection direction $\widehat{u}$ constructed in \eqref{eq: constraint 1} and \eqref{eq: constraint 2} enables the effective bias correction for any loading vector $\xnew$, no matter it is sparse or dense. The work \cite{cai2021optimal} established the asymptotic normality of the estimator $\widehat{x_{\rm new}^{\intercal}\beta}$ in \eqref{eq: LF linear} for any loading vector $\xnew\in \bbR^{p+1}.$
Based on the asymptotic normality, we construct a confidence interval for $\xnew^{\intercal}\beta$ as
\begin{equation}
\small
	\mathrm{CI}=\left(\widehat{\xnew^{\intercal}\beta}-z_{\alpha / 2} \sqrt{\widehat{\mathrm{V}}}, \quad \widehat{\xnew^{\intercal}\beta}+z_{\alpha / 2} \sqrt{\widehat{\mathrm{V}}}\right) \quad \text{with}\quad \widehat{\mathrm{V}}=\frac{\widehat{\sigma}_{\epsilon}^{2}}{n} \widehat{u}^{\intercal} \widehat{\Sigma} \widehat{u}.
	\label{eq: CI linear LF}
\end{equation}

%\begin{Remark}[Plug-in Debiased Estimators] \rm A natural way of making inferences for $\xnew^{\intercal}\beta$ is to plug in the debiased estimators $\widehat{\beta}^{\rm Deb}_j$ defined in \eqref{eq: debiased estimator 0}, that is, $\sum_{j=1}^{p+1} [\xnew]_j \widehat{\beta}^{\rm Deb}_j.$ There exist two problems with this plug-in debiased estimator: firstly, the accumulation of the remaining bias components of $\{\widehat{\beta}^{\rm Deb}_j\}_{1\leq j\leq p+1}$ might lead to a large bias component of this plug-in estimator; %\Yin{ I am not sure if the previous wording is rigorous since the first component is not regularized?} 
%	secondly, it is time-consuming to debias each regression coefficient as this requires implementing $p$ extra high-dimensional optimization problems. 
%\end{Remark}

%%%%%%%%%%%%%%%%%%%%%%%%%%%%
%%%%%%%%%%%%%%%%%%%%%%%%%%%%
\subsection{Linear functionals for logistic regression}
\label{sec: LF logistic}
%%%%%%%%%%%%%%%%%%%%%%%%%%%%
%%%%%%%%%%%%%%%%%%%%%%%%%%%%
We now consider the high-dimensional logistic model and present the inference procedures for $\xnew^{\intercal}\beta$ or $h(\xnew^{\intercal}\beta)$ proposed in \cite{guo2021inference}. In particular, \cite{guo2021inference} proposed the following debiased estimator, 
\begin{equation}
\small
	\widehat{\xnew^{\intercal}{\beta}}=\xnew^{\intercal}\widehat{\beta}+\widehat{u}^{\intercal}\frac{1}{n}\sum_{k=1}^{n}W_kX_{k,\cdot}\left(Y_k-h(X_{k,\cdot}^{\intercal}\widehat{\beta})\right),
	\label{eq: bias correction general}
\end{equation}
where $\widehat{\beta}$ is defined in \eqref{eq: penalized MLE}, $W_{k}=1/h'(X_{k,\cdot}^{\intercal}\widehat{\beta})$ for $1 \leq k \leq n$, and the projection direction $\widehat{u} \in\bbR^{p+1}$ is defined as
\begin{equation}
\small
	\begin{aligned}
		\widehat{u}&=\;\argmin_{u\in \bbR^{p+1}} u^{\intercal} \left(\frac{1}{n}\sum_{k=1}^{n} W_k h'(X_{k,\cdot}^{\intercal}\widehat{\beta})X_{k,\cdot}X_{k,\cdot}^{\intercal}\right) u\\
		\text{subject to}&\quad \left \|\left(\frac{1}{n}\sum_{k=1}^{n} W_k h'(X_{k,\cdot}^{\intercal}\widehat{\beta})X_{k,\cdot}X_{k,\cdot}^{\intercal}\right) u-\xnew\right\|_{\infty}{\leq \|x_{\rm new}\|_2\lambda}, \quad \|Xu\|_{\infty}\leq \tau\\
		&\quad \left\vert x_{\rm new}^{\intercal} \left(\frac{1}{n}\sum_{k=1}^{n} W_k h'(X_{k,\cdot}^{\intercal}\widehat{\beta})X_{k,\cdot}X_{k,\cdot}^{\intercal}\right) u-\|x_{\rm new}\|_2^2 \right\vert\leq \|x_{\rm new}\|_2^2\lambda. 
	\end{aligned}
	\label{eq: weighted projection}
\end{equation}
The bias-corrected estimator in \eqref{eq: bias correction general} can be viewed as a generalization of those in \eqref{eq: constraint 1} and \eqref{eq: constraint 2} by incorporating the weighted bias-correction in \eqref{eq: generic form}.

It has been established in \cite{guo2021inference} that $\widehat{\xnew^{\intercal}{\beta}}$ in \eqref{eq: bias correction general} is asymptotically unbiased and normal.
Based on the asymptotic normality, we construct the following confidence interval 
\begin{equation}
\small
	\mathrm{CI}=\left(\widehat{\xnew^{\intercal} \beta}-z_{\alpha / 2} \sqrt{\widehat{\mathrm{V}}}, \quad \widehat{\xnew^{\intercal} \beta}+z_{\alpha / 2} \sqrt{\widehat{\mathrm{V}}}\right) \quad \text{with}\quad \widehat{\mathrm{V}}=\frac{1}{n} \widehat{u}^{\intercal} \widehat{\Sigma}^{G} \widehat{u},
	\label{eq: CI logistic LF}
\end{equation}
where $\widehat{\Sigma}^{G}=\frac{1}{n}\sum_{k=1}^{n} W_k^2h(X_{k,\cdot}^{\intercal}\widehat{\beta})(1-h(X_{k,\cdot}^{\intercal}\widehat{\beta}))X_{k,\cdot} X_{k,\cdot}^{\intercal}$.  We estimate the case probability $\PP(Y_k=1\vert X_{k,\cdot}=\xnew)$ by $h(\widehat{\xnew^{\intercal}\beta})$ and construct the confidence interval for $h(x_{\text{new}}^{\intercal}\beta)$ as
\begin{equation}
\small
	{\rm CI}=\left[h\left(\widehat{\xnew^{\intercal} \beta}-z_{\alpha/2}\sqrt{\widehat{\mathrm{V}}}\right),h\left(\widehat{\xnew^{\intercal} \beta}+z_{\alpha/2}\sqrt{\widehat{\mathrm{V}}}\right)\right].
	\label{eq: CI logistic LF}
\end{equation}

%%%%%%%%%%%%%%%%%%%%%%%%%%%%%%%%%%%%%%%%%%%%%%%%%%%%%%%%%%%%%%%%%%%%%%%%%%%%%%%%%%%%%%%%%%%%%%%%%%%%%%%%%%%
\subsection{Conditional average treatment effects}
%%%%%%%%%%%%%%%%%%%%%%%%%%%%%%%%%%%%%%%%%%%%%%%%%%%%%%%%%%%%%%%%%%%%%%%%%%%%%%%%%%%%%%%%%%%%%%%%%%%%%%%%%%%

The inference methods proposed in Sections \ref{sec: LF linear} and \ref{sec: LF logistic} can be generalized to make inferences for conditional average treatment effects, which can be expressed as the difference between two linear functionals. For $1\leq k\leq n,$ let $A_k\in \{1,2\}$ denote the treatment assignment for the $k^{\th}$ observation, where $A_k=1$ and $A_k=2$ represent the subject receiving the control or the treatment assignment, respectively. Moreover, in the context of comparing the treatment effectiveness, $A_k=1$, and $A_k=2$ may stand for the subject receiving the first or the second treatment assignment, respectively. 
As a special case of \eqref{eq: multi-regression}, we consider the following conditional outcome models 
$\bbE (Y_k\vert X_{k,\cdot}, A_k=1)= X_{k,\cdot}^{\intercal}\beta^{\one}$ and $\bbE (Y_k\vert X_{k,\cdot}, A_k=2)= X_{k,\cdot}^{\intercal}\beta^{\two}.$ For an individual with $X_{k,\cdot}=\xnew$, we define
$
\Delta(\xnew)=\bbE (Y_k\vert X_{k,\cdot}=\xnew, A_k=2)-\bbE (Y_k\vert X_{k,\cdot}=\xnew, A_k=1)=\xnew^{\intercal}(\beta^{\two}-\beta^{\one}),
$ which measures the change of the conditional mean from being untreated to treated for individuals with the covariates $\xnew$. 

By generalizing \eqref{eq: LF linear}, we construct the debiased estimators $\widehat{\xnew^{\intercal}\beta^{\one}}$ and $\widehat{\xnew^{\intercal}\beta^{\two}},$ together with their corresponding variance estimators $\widehat{\mathrm{V}}_{\beta^{\one}}$ and $\widehat{\mathrm{V}}_{\beta^{\two}}$. The paper \cite{cai2021optimal} proposed to estimate $\Delta(\xnew)$ by $\widehat{\Delta}(\xnew)=\widehat{\xnew^{\intercal}\beta^{\two}}-\widehat{\xnew^{\intercal}\beta^{\one}}$ and construct the confidence interval for $\Delta(\xnew)$ as 
\begin{equation}
\small
	\mathrm{CI}=\left(\widehat{\Delta}(\xnew)-z_{\frac{\alpha}{2}} \sqrt{\widehat{\mathrm{V}}_{\beta^{\one}}+\widehat{\mathrm{V}}_{\beta^{\two}}}, \quad \widehat{\Delta}(\xnew)+z_{\frac{\alpha}{2}} \sqrt{\widehat{\mathrm{V}}_{\beta^{\one}}+\widehat{\mathrm{V}}_{\beta^{\two}}}\right).
	\label{eq: CI linear CATE}
\end{equation}
Regarding the hypothesis testing problem
$
H_{0}: \xnew^{\intercal}(\beta^{\two}-\beta^{\one}) \leq 0$ versus $H_{1}: \xnew^{\intercal}(\beta^{\two}-\beta^{\one})>0,$
the paper \cite{cai2021optimal} proposed the following test procedure,
\begin{equation}
\small
	{\phi}_{\alpha}(\xnew)=\bI\left\{\widehat{\Delta}(\xnew)-z_{\alpha} \sqrt{\widehat{\mathrm{V}}_{\beta^{\one}}+\widehat{\mathrm{V}}_{\beta^{\two}}}\geq 0\right\}.
	\label{eq: test logistic}
\end{equation}

As a direct generalization, we may consider the logistic version,
$
\bbE (Y_k\vert X_{k,\cdot}, A_k=1)= h(X_{k,\cdot}^{\intercal}\beta^{\one})$ and $\bbE (Y_k\vert X_{k,\cdot}, A_k=2)= h(X_{k,\cdot}^{\intercal}\beta^{\two}).$ For an individual with $X_{k,\cdot}=\xnew$, our inference target becomes $
\Delta(\xnew)=h(\xnew^{\intercal}\beta^{\two})-h(\xnew^{\intercal}\beta^{\one}).$ The methods in Section \ref{sec: LF logistic} can be applied here to make inferences for $\Delta(\xnew).$

%%%%%%%%%%%%%%%%%%%%%%%%%%%%
%%%%%%%%%%%%%%%%%%%%%%%%%%%%
\subsection{Quadratic functionals}
\label{sec: QF}
%%%%%%%%%%%%%%%%%%%%%%%%%%%%
%%%%%%%%%%%%%%%%%%%%%%%%%%%%
We now focus on inference for the quadratic functionals $\QA=\beta^{\intercal}_{G}A \beta_{G}$ and $\QW=\beta_{G}^{\intercal}\Sigma_{G,G}\beta_{G}$, where $G\subset\{1,\cdots,p+1\}$ and $A\in \bbR^{\vert G\vert\times \vert G\vert}$ denotes a pre-specified matrix. Without loss of generality, we set $G=\{2,\cdots, \vert G\vert+1\}.$   
In the following, we mainly discuss the main idea under the high-dimensional linear regression, which can be generalized to the high-dimensional logistic regression. Let $\betainit$ denote the Lasso estimator in \eqref{eq: Lasso}. 

We start with the error decomposition of the plug-in estimator $\betainit_{\setG}^{\intercal}A\betainit_{\setG}$,
\begin{equation}
\small
	\betainit_{\setG}^{\intercal}A\betainit_{\setG}-\beta_{\setG}^{\intercal}A{\beta_{\setG}}=2\betainit_{\setG}^{\intercal}A(\betainit_{\setG}-\beta_{\setG})
	-(\betainit_{\setG}-\beta_{\setG})^{\intercal}A(\betainit_{\setG}-\beta_{\setG}).
	\label{eq: key decomp QF}
\end{equation}
In consideration of high-dimensional linear models, 
\cite{guo2021group,guo2019optimal} proposed to construct the bias-corrected estimator through estimating the error component $2\betainit_{\setG}^{\intercal}A(\betainit_{\setG}-\beta_{\setG})$ on the right-hand side of \eqref{eq: key decomp QF}. Since $\betainit_{\setG}^{\intercal}A(\betainit_{\setG}-\beta_{\setG})$ can be expressed as $\xnew^{\intercal}(\widehat{\beta}-\beta)$ with $\xnew=\begin{pmatrix} 0&\betainit_{\setG}^{\intercal}A& {\bf 0}^{\intercal}\end{pmatrix}^{\intercal}$, the techniques of estimating the error component for the linear functional can be directly applied to approximate $\betainit_{\setG}^{\intercal}A(\betainit_{\setG}-\beta_{\setG})$. Particularly, \cite{guo2021group,guo2019optimal} proposed the following estimator of $\QA$, 
\begin{equation}
\small
	\QAh=\betainit_{\setG}^{\intercal}A\betainit_{\setG}+2\widehat{u}_{A}^{\intercal}X^{\intercal}(Y-X\betainit)/n,
	\label{eq: general A}
\end{equation}
where $\widehat{u}_{A}$ denotes the solution of \eqref{eq: constraint 1} and \eqref{eq: constraint 2} with $\xnew=\begin{pmatrix} 0&\betainit_{\setG}^{\intercal}A& {\bf 0}^{\intercal}\end{pmatrix}^{\intercal}.$ No bias correction is required for the last term on the right-hand side of \eqref{eq: key decomp QF} since it is a higher-order error term under regular conditions.

We now turn to the estimation of $\QW=\beta_{G}^{\intercal}\Sigma_{G,G}\beta_{G},$ where the matrix $\Sigma_{G,G}$ has to be estimated from the data. 
With $\widehat{\Sigma}=\frac{1}{n}\sum_{k=1}^{n}X_{k,\cdot} X_{k,\cdot}^{\intercal},$ we decompose the estimation error of the plug-in estimator $\betainit^{\intercal}_{G} \widehat{\Sigma}_{G,G}\betainit_{G}$ as
\begin{equation}
\small
	\begin{aligned}
	\betainit^{\intercal}_{G} \widehat{\Sigma}_{G,G}\betainit_{G}-\beta_{G}^{\intercal}\Sigma_{G,G} \beta_{G}=&2\betainit_{G}^{\intercal}\widehat{\Sigma}_{G,G}(\betainit_{G}-\beta_{G})+\beta_{G}^{\intercal}(\widehat{\Sigma}_{G,G}-\Sigma_{G,G})\beta_{G}\\ 
	&-(\betainit_{G}-\beta_{G})^{\intercal}\widehat{\Sigma}_{G,G}(\betainit_{G}-\beta_{G}).
	\end{aligned}
	\label{eq: key decomp QF Sigma}
\end{equation}
By a similar approach as \eqref{eq: general A}, \cite{guo2021group} proposed the following estimator of $\QW$, 
\begin{equation}
\small
	\QWh=\betainit^{\intercal}_{G} \widehat{\Sigma}_{G,G}\betainit_{G}+2\widehat{u}_{\Sigma}^{\intercal}X^{\intercal}(Y-X\betainit)/n,
	\label{eq: general Sigma}
\end{equation}
where $\widehat{u}_{\Sigma}$ denotes the solution of \eqref{eq: constraint 1} and \eqref{eq: constraint 2} with $\xnew=\begin{pmatrix} 0&\betainit_{\setG}^{\intercal}\widehat{\Sigma}_{G,G}& {\bf 0}^{\intercal}\end{pmatrix}^{\intercal}.$ As a special case, \cite{cai2020semisupervised} considered $G=\{1,\cdots,p+1\}$ and constructed $\widehat{u}_{\Sigma}=\widehat{\beta}.$

The works \cite{guo2021group,guo2019optimal} established the asymptotic properties for a sample-splitting version of $\QAh$ and $\QWh$, where the initial estimator $\betainit$ is constructed from a subsample independent from the samples $\{X_{k,\cdot},Y_k\}_{1\leq k\leq n}$ used in \eqref{eq: general A} and \eqref{eq: general Sigma}. Based on the asymptotic properties, \cite{guo2021group} estimated the variance of $\QAh$ by 
$
\VAh={4\widehat{\sigma}_\epsilon^2}\widehat{u}_{A}^{\intercal}\widehat{\Sigma}\widehat{u}_{A}/n+{\tau}/{n}
$ 
for some positive constant $\tau>0$, where the term $\tau/n$ is introduced as an upper bound for the term $(\betainit_{\setG}-\beta_{\setG})^{\intercal}A(\betainit_{\setG}-\beta_{\setG})$ in \eqref{eq: key decomp QF}. \cite{guo2021group} estimated the variance of $\QWh$ by
$$\small\VWh=\frac{4 \widehat{\sigma}_{\epsilon}^{2}}{n} \widehat{u}_{\Sigma}^{\intercal} \widehat{\Sigma} \widehat{u}_{\Sigma}+
\frac{1}{n^{2}} \sum_{k=1}^{n}\left(\widehat{\beta}_{G}^{\intercal} X_{k, G} X_{k, G}^{\intercal} \widehat{\beta}_{G}-\widehat{\beta}_{G}^{\intercal} \widehat{\Sigma}_{G, G} \widehat{\beta}_{G}\right)^{2}
+\frac{\tau}{n},$$
and further constructed the following confidence intervals for $\QA$ and $\QW$,  
\begin{equation}
\small
	\begin{aligned}
		{\rm CI}_{A}(\tau)&=\left(\QAh-{z_{\alpha/2}}\cdot \sqrt{\VAh}, \,\, \QAh+{z_{\alpha/2}} \cdot \sqrt{\VAh}\right),\\
		{\rm CI}_{\Sigma}(\tau)&=\left(\QWh-{z_{\alpha/2}}\cdot \sqrt{\VWh}, \,\, \QWh+{z_{\alpha/2}} \cdot \sqrt{\VWh}\right).
	\end{aligned}
	\label{eq: CI QF linear}
\end{equation}
%where $z_{\alpha/2}$ is the upper $\alpha/2$ quantile for the standard normal distribution. 

For a positive definite matrix $A$, the significance test $H_0: \beta_G = 0$ can be recast as 
$H_{0, A}: \beta_{G}^{\intercal} A \beta_{G}=0$ or $H_{0, \Sigma}: \beta_{G}^{\intercal} \Sigma_{G,G} \beta_{G}=0.$
The following $\alpha$-level significance tests of $H_{0, A}$ and $H_{0,\Sigma}$ have been respectively proposed in \cite{guo2021group}, 
\begin{equation}
\small
	\phi_{A}(\tau)=\bI\left\{\QAh\geq z_{\alpha} \cdot \sqrt{{\VAh}} \right\},\quad \phi_{\Sigma}(\tau)=\bI\left\{\QWh\geq {z_{\alpha}} \cdot \sqrt{\VWh} \right\}.
	\label{eq: test QF linear}
\end{equation}

%with $z_{\alpha}$ denoting the upper $\alpha$ quantile for the standard normal distribution. 

The inference for $\QA$ and $\QW$ can be generalized to the high-dimensional logistic regression together with the weighted bias correction detailed in \eqref{eq: weighted projection}. The paper \cite{ma2022statistical} studied how to construct the debiased estimators of $\QW$ under the high-dimensional logistic regression. 

%\begin{Remark}[Plug-in Debiased Estimators] \rm We use inference for $\QA$ as an example to explain that plugging in the debiased estimator $\widehat{\beta}^{\rm Deb}_j$ in \eqref{eq: debiased estimator 0} leads to variance inflation. When the group $G$ is large, the plug-in estimator $\sum_{i,j\in G} A_{i,j} \widehat{\beta}^{\rm Deb}_i\widehat{\beta}^{\rm Deb}_j$ suffers from a large variance since the addition of uncertainty of $\{\widehat{\beta}^{\rm Deb}_j\}_{j \in G}$ can significantly inflate the variance of the plug-in estimator for a large group $G$. In contrast, the debiased estimator $\QAh$ in \eqref{eq: general A} balances the bias and variance and achieves the optimal convergence rate, as discussed in the following subsection. 
%\end{Remark}

%%%%%%%%%%%%%%%%%%%%%%%%%%%%
\subsection{Semi-supervised inference} %optimal estimation of $\|\beta\|_2^2$ and $\beta^{\intercal}\Sigma\beta$}
\label{sec: semi-super}
%%%%%%%%%%%%%%%%%%%%%%%%%%%%
We summarize the optimal estimation of $\|\beta\|_2^2$ and $\beta^{\intercal}\Sigma\beta$ in a general semi-supervised setting, with the labelled data $(X_{1,\cdot},Y_1),\cdots, (X_{n,\cdot},Y_n)$ and the unlabelled data $X_{n+1,\cdot},\cdots, X_{n+N,\cdot}$, where the covariates $X_{1,\cdot},\cdots, X_{n+N,\cdot}$ are assumed to be identically distributed.  
%The information of $\Sigma$ is critical for a collection of statistical inference problems. The papers \cite{javanmard2018debiasing,cai2017confidence} demonstrated how the knowledge of $\Sigma$ can facilitate the inference for $\beta_j$ under the high-dimensional linear model. In the following, we focus on inference for $\beta^{\intercal}\Sigma\beta$ and $\|\beta\|_2^2$ in the semi-supervised setting. 
%We consider the semi-supervised setting 

The work \cite{cai2020semisupervised} proposed the following semi-supervised estimator of $\beta^{\intercal}\Sigma\beta$, 
\begin{equation}
\small
	{\widehat{\rmQ}}(\widehat{\beta},\widehat{\Sigma}^{S})=\widehat{\beta}^{\intercal} \widehat{\Sigma}^{S}\widehat{\beta}+2\widehat{\beta}^{\intercal}\frac{1}{n}\sum_{k=1}^{n}X_{k,\cdot}(Y_k-X_{k,\cdot}^{\intercal}\widehat{\beta}),
	 \widehat{\Sigma}^{S}=\frac{1}{n+N}\sum_{k=1}^{n+N}X_{k,\cdot}X_{k,\cdot}^{\intercal}.
	\label{eq: EST semi}
\end{equation}
The matrix estimator $\widehat{\Sigma}^{S}$ in \eqref{eq: EST semi} utilizes both the labeled and unlabelled data. 
%This generalizes the supervised estimator in \eqref{eq: general Sigma} with $G=\{1,\cdots,p+1\}$ and $\widehat{u}_{\Sigma}=\widehat{\beta}$ by incorporating the additional information contained in the unlabelled data. In addition, 
The work \cite{cai2020semisupervised} constructed the following confidence interval for $\beta^{\intercal}\Sigma\beta$ in the semi-supervised setting,
\begin{equation*}
\small
{\rm CI}(Z)=\left(\left({\widehat{\rmQ}}(\widehat{\beta},\widehat{\Sigma}^{S})-z_{\alpha/2}\sqrt{\widehat{\rm V}}\right)_{+},\; {\widehat{\rmQ}}(\widehat{\beta},\widehat{\Sigma}^{S})-z_{\alpha/2}\sqrt{\widehat{\rm V}}\right),
\end{equation*}
where $\small \widehat{\rm V}=\frac{1}{n}\left(4 {\widehat{\sigma}_\epsilon^2\widehat{\beta}^{\intercal}\widehat{\Sigma}^{S}\widehat{\beta}}+{\frac{\widehat{\rho}}{n+N}\sum_{ k=1}^{n+N} \left(\widehat{\beta}^{\intercal} X_{k\cdot} X_{k\cdot}^{\intercal}\widehat{\beta}-\widehat{\beta}^{\intercal}\widehat{\Sigma}^{S}\widehat{\beta}\right)^2}+\tau\right)$ with $\widehat{\rho}=n/(N+n)$ and $\tau>0$ being a user-specific tuning parameter adjusting for the higher order estimation error. The above confidence interval construction demonstrates the usefulness of integrating the unlabelled data, significantly reducing the ratio $\widehat{\rho}$ and the interval length. 

Define $\Theta\left(s,M\right)=\left\{\left(\beta,\Sigma,\sigma_\epsilon\right)\in \Theta(s):
{\frac{M}{2}\leq \|\beta\|_2\leq M} \right\},$ as a subspace of $\Theta(s)$ in \eqref{eq: parameter space} with the constraint on $\|\beta\|_2.$ In the semi-supervised setting, it was established in \cite{cai2020semisupervised} that the minimax rate for estimating $\beta^{\intercal}{\Sigma}\beta$ over $\Theta\left(s,M\right)$ is 
\begin{equation}
\small
	{M^2\over \sqrt{N+n}}+ \min\left\{{M\over \sqrt{n}} + {s \log p\over n}, M^2\right\}.
	\label{eq: semi optimal}
\end{equation}
The estimator ${\widehat{\rmQ}}(\widehat{\beta},\widehat{\Sigma}^{S})$ in \eqref{eq: EST semi} is shown to achieve the optimal rate in \eqref{eq: semi optimal} for $M\geq C\sqrt{s\log p/n}$. When $M\leq C\sqrt{s\log p/n}$, then estimating $\rmQ$ by zero achieves the optimal rate in \eqref{eq: semi optimal}. The optimal convergence rate in \eqref{eq: EST semi} characterizes its dependence on the amount of unlabelled data. With a larger size $N$ of the unlabelled data, the convergence rate ${M^2\over \sqrt{N+n}}$ decreases. In the semi-supervised setting, the unlabelled data can also be used to construct a more accurate estimator of $\|\beta\|_2^2$; see Section 4 of \cite{cai2020semisupervised} for details. 

In the following, we consider the two extremes under semi-supervised learning: the supervised learning setting with $N=0$ and the other extreme setting of knowing $\Sigma={\rm I}$, which can be viewed as a special case of the semi-supervised setting with $N\rightarrow \infty.$ In Table \ref{tab: optimal}, we summarize the optimal convergence rate of estimating $\|\beta\|_2^2$ and $\beta^{\intercal}\Sigma\beta$ over both $\Theta(s,M)$ and $\Theta_0(s,M)$, where $\Theta_0(s,M)=\left\{\left(\beta,\Sigma,\sigma_\epsilon\right)\in \Theta(s,M):
\Sigma={\rm I} \right\}.$ With leveraging the information of $\Sigma$, Table \ref{tab: optimal} shows that the optimal rates of estimating $\|\beta\|_2^2$ and $\beta^{\intercal}\Sigma\beta$ are reduced by ${M\frac{s \log p}{n}}$ and ${M^2\frac{1}{\sqrt{n}}},$ respectively.  The improvement can be substantial when the parameter $M$, characterizing the $\ell_2$ norm $\|\beta\|_2$, is relatively large. The optimal rate of estimating $\|\beta\|_2^2$ was established in \cite{collier2017minimax} under the sequence model $Y_j=\beta_j+ \frac{1}{\sqrt{n}}\epsilon_j \; \text{for}\; 1\leq j\leq p.$ When we know $\Sigma={\rm I}$ and focus on the regime $s\leq c \min\{{n/\log p}, p^{\nu}\}$ for $0\leq \nu<1/2$, the optimal rate of estimating $\|\beta\|_2^2$ in the high-dimensional linear model matches that in \cite{collier2017minimax}. However, when $\Sigma$ is unknown,  estimating $\|\beta\|_2^2$ is much harder in the high-dimensional linear model than in the sequence model.
 %Note that the parameter space $\Theta_0(s,M)$ corresponds to the extreme setting with the knowledge $\Sigma={\rm I}.$

\begin{table}[h]
	\centering
	\renewcommand{\arraystretch}{1.75}
	\begin{tabular}{|c|c|c|c|}
		\hline
		Target & Optimal Rate over $\Theta(s,M)$ &Optimal Rate over $\Theta_0(s,M)$\\ 
		\hline
		$\|\beta\|_2^2$& $\min\left\{M\frac{1}{\sqrt{n}}+ \frac{s \log p}{n}+{M\frac{s \log p}{n}}, M^2\right\}$ & \multirow{2}{*}{$\min\left\{M\frac{1}{\sqrt{n}}+ \frac{s \log p}{n}, M^2\right\}$} \\
		$\beta^{\intercal}\Sigma\beta$ & $\min\left\{M\frac{1}{\sqrt{n}}+ \frac{s \log p}{n}+{M^2\frac{1}{\sqrt{n}}}, M^2\right\}$ & \\
		\hline
	\end{tabular}
	\caption{The minimax optimal rate of estimating $\|\beta\|_2^2$ over $\Theta(s,M)$ was established in \cite{guo2019optimal}; the remaining optimal rates are implied by \eqref{eq: semi optimal}, which was established in \cite{cai2020semisupervised}. The focus is on the regime $s\leq c \min\{{n/\log p}, p^{\nu}\}$ for $0\leq \nu<1/2.$}
	\label{tab: optimal}
\end{table}

\vspace{-7mm}

%%%%%%%%%%%%%%%%%%%%%%%%%%%%
%%%%%%%%%%%%%%%%%%%%%%%%%%%%
\subsection{Inner products of regression vectors}
\label{sec: inner}
%%%%%%%%%%%%%%%%%%%%%%%%%%%%
%%%%%%%%%%%%%%%%%%%%%%%%%%%%
%\begin{equation}
%\begin{aligned}
%\E(Y^{\d}_k\vert X^{\d}_{k,\cdot})=h([X^{\d}_{k,\cdot}]^{\intercal}\beta^{\d}) \quad \text{for}\quad 1\leq k\leq n_1, %\quad \E(W_j\vert Z_{j,\cdot}) = h(Z_{j,\cdot}^{\intercal}\gamma) \quad \text{for}\quad1\leq j \leq n_2,
%\end{aligned}
%\label{eq: multi-regression}
%\end{equation}
We next consider the two-sample regression models in \eqref{eq: multi-regression}.  %and define the regression errors $\epsilon^{\d}_k=Y^{\d}_k-h([X^{\d}_{k,\cdot}]^{\intercal}\beta^{\d})$ for $1\leq k\leq n_d$, $d=1,2.$
%\eqref{eq: multi-regression} as \begin{equation}
	%\begin{aligned}
	%Y_k = X_{k,\cdot}^{\intercal}\beta+\epsilon_k \quad \text{for}\quad 1\leq k\leq n_1, \quad W_k = Z_{j,\cdot}^{\intercal}\gamma+\delta_i \quad \text{for}\quad1\leq k\leq n_2,
	%\end{aligned}
	%\label{eq: multi-regression rewrite}
	%\end{equation}
	%where $X_{k,\cdot}\in \R^{p+1}$ with $1\leq k\leq n_1$ and $Z_{j,\cdot}\in \R^{p+1}$ with $1\leq k\leq n_2$ denote the high-dimensional covariates. 
When the high-dimensional covariates are genetic variants and the outcome variables measure different phenotypes, then the inner product $[\beta^{\one}]^{\intercal}\beta^{\two}$ can be interpreted as the genetic relatedness \cite{guo2019optimal}, measuring the similarity between the two association vectors $\beta^{\one}$ and $\beta^{\two}$. We present the debiased estimator of $[\beta^{\one}]^{\intercal}\beta^{\two}$ proposed in \cite{guo2019optimal}. For $d=1,2,$ let $\widehat{\beta}^{\ssd}$ be the Lasso estimator of $\beta^{\ssd}$. Similar to the decomposition in \eqref{eq: key decomp QF}, we decompose the error of the plug-in estimator $[\widehat{\beta}^{\one}]^{\intercal}\widehat{\beta}^{\two},$
	\begin{equation}
	\small
		\begin{aligned}
		[\widehat{\beta}^{\one}]^{\intercal}\widehat{\beta}^{\two}-[\beta^{\one}]^{\intercal}\beta^{\two}=&[\widehat{\beta}^{\one}]^{\intercal}(\widehat{\beta}^{\two}-\beta^{\two})+[\widehat{\beta}^{\two}]^{\intercal}(\widehat{\beta}^{\one}-\beta^{\one})\\
		&-
		(\widehat{\beta}^{\two}-\beta^{\two})^{\intercal}(\widehat{\beta}^{\one}-\beta^{\one}).
		\end{aligned}
		\label{eq: key decomp Inner}
	\end{equation}
The key is to estimate $[\widehat{\beta}^{\one}]^{\intercal}(\widehat{\beta}^{\two}-\beta^{\two})$ and $[\widehat{\beta}^{\two}]^{\intercal}(\widehat{\beta}^{\one}-\beta^{\one})$ separately in the above decomposition, which can be viewed as a projection of $\widehat{\beta}^{\two}-\beta^{\two}$ and $\widehat{\beta}^{\one}-\beta^{\one},$ respectively.  

The work \cite{guo2019optimal} proposed the following bias-corrected estimator of $[\beta^{\one}]^{\intercal}\beta^{\two},$
	\begin{equation}
	\small
		\begin{aligned}
		\widehat{[\beta^{\one}]^{\intercal}\beta^{\two}}=&[\widehat{\beta}^{\one}]^{\intercal}\widehat{\beta}^{\two}+\widehat{u}_1^{\intercal}\frac{1}{n_2} \sum_{k=1}^{n_2}[X^{\two}_{k,\cdot}]^{\intercal}(Y^{\two}_k-[X^{\two}_{k,\cdot}]^{\intercal}\widehat{\beta}^{\two})\\
		&+\widehat{u}_2^{\intercal}\frac{1}{n_1} \sum_{k=1}^{n_1}[X^{\one}_{k,\cdot}]^{\intercal}(Y^{\one}_k-[X^{\one}_{k,\cdot}]^{\intercal}\widehat{\beta}^{\one})
		\end{aligned}
		\label{eq: debiased inner}
	\end{equation}
	where the projection direction vectors are constructed as
	\begin{equation*}
	\small
		\begin{aligned}
			\widehat{u}_1=\;\argmin_{u\in \bbR^{p+1}} u^{\intercal} \widehat{\Sigma}^{\two} u \quad \text{subject to}\;
			& \left \|\widehat{\Sigma}^{\two} u-\widehat{\beta}^{\one}\right\|_{\infty}\leq \|\widehat{\beta}^{\one}\|_2 \lambda_2 \\
			\widehat{u}_2=\;\argmin_{u\in \bbR^{p+1}} u^{\intercal} \widehat{\Sigma}^{\one} u \quad \text{subject to}\;
			& \left \|\widehat{\Sigma}^{\one} u-\widehat{\beta}^{\two}\right\|_{\infty}\leq \|\widehat{\beta}^{\two}\|_2 \lambda_1
		\end{aligned}
	\end{equation*}
	with $\widehat{\Sigma}^{\ssd}=\frac{1}{n}\sum_{k=1}^{n_d}X^{\ssd}_{k,\cdot} [X^{\ssd}_{k,\cdot}]^{\intercal}$ and $\lambda_d\asymp \sqrt{{\log p}/{n_d}}$ for $d=1,2.$ 
		
	For the estimator \eqref{eq: debiased inner}, the bias-correction terms $\widehat{u}_1^{\intercal}\frac{1}{n_2} \sum_{k=1}^{n_2}[X^{\two}_{k,\cdot}]^{\intercal}(Y^{\two}_k-[X^{\two}_{k,\cdot}]^{\intercal}\widehat{\beta}^{\two})$ and $\widehat{u}_2^{\intercal}\frac{1}{n_1} \sum_{k=1}^{n_1}[X^{\one}_{k,\cdot}]^{\intercal}(Y^{\one}_k-[X^{\one}_{k,\cdot}]^{\intercal}\widehat{\beta}^{\one})$ are constructed to estimate $[\widehat{\beta}^{\one}]^{\intercal}({\beta}^{\two}-\widehat{\beta}^{\two})$ and $[\widehat{\beta}^{\two}]^{\intercal}({\beta}^{\one}-\widehat{\beta}^{\one}),$ respectively.  
	The construction of the projection directions $\widehat{u}_1$ and $\widehat{u}_2$ can be viewed as extensions of that in \eqref{eq: constraint 1} with $\xnew=\widehat{\beta}^{\one}$ and $\xnew=\widehat{\beta}^{\two},$ respectively. An additional constraint as in \eqref{eq: constraint 2} is not needed here since both $\xnew=\widehat{\beta}^{\one}$ and $\xnew=\widehat{\beta}^{\two}$ are sufficiently sparse. The paper \cite{guo2019optimal} has established the convergence rate of the debiased estimator proposed in \eqref{eq: debiased inner}. The analysis can be extended to establish the asymptotic normal distribution of $\widehat{[\beta^{\one}]^{\intercal}\beta^{\two}}$ under suitable conditions.  
	
The quantity $[\beta^{\one}]^{\intercal}\Sigma\beta^{\two}$ is another genetic relatedness measure if $\Sigma^{\one}=\Sigma^{\two}=\Sigma$ with $\Sigma^{\ssd}=\bbE X^{\ssd}_{k,\cdot} [X^{\ssd}_{k,\cdot}]^{\intercal}$ for $d=1,2.$ We can propose the debiased estimator $\widehat{[\beta^{\one}]^{\intercal}\Sigma\beta^{\two}}$, defined as $[\widehat{\beta}^{\one}]^{\intercal}\widehat{\Sigma}\widehat{\beta}^{\two}+[\widehat{\beta}^{\one}]^{\intercal}\frac{1}{n_2} \sum_{k=1}^{n_2}[X^{\two}_{k,\cdot}]^{\intercal}(Y^{\two}_k-[X^{\two}_{k,\cdot}]^{\intercal}\widehat{\beta}^{\two})+[\widehat{\beta}^{\two}]^{\intercal}\frac{1}{n_1} \sum_{k=1}^{n_1}[X^{\one}_{k,\cdot}]^{\intercal}(Y^{\one}_k-[X^{\one}_{k,\cdot}]^{\intercal}\widehat{\beta}^{\one}).
$
%	\begin{equation*}
%	\small
%		\begin{aligned}
%		&[\widehat{\beta}^{\one}]^{\intercal}\widehat{\Sigma}\widehat{\beta}^{\two}+[\widehat{\beta}^{\one}]^{\intercal}\frac{1}{n_2} \sum_{k=1}^{n_2}[X^{\two}_{k,\cdot}]^{\intercal}(Y^{\two}_k-[X^{\two}_{k,\cdot}]^{\intercal}\widehat{\beta}^{\two})\\
%		&+[\widehat{\beta}^{\two}]^{\intercal}\frac{1}{n_1} \sum_{k=1}^{n_1}[X^{\one}_{k,\cdot}]^{\intercal}(Y^{\one}_k-[X^{\one}_{k,\cdot}]^{\intercal}\widehat{\beta}^{\one}).
%		\end{aligned}
%		%\label{eq: debiased linear}
%	\end{equation*}
%	This estimator also generalizes the estimator of the quadratic norm $\beta^{\intercal}\Sigma\beta$ in one-sample setting proposed in \cite{cai2020semisupervised}. 
	The above results have been extended to the logistic regression models, where the quantity $[\beta^{\one}]^{\intercal}\Sigma\beta^{\two}$ still captures an interpretation of genetic relatedness. The paper \cite{ma2022statistical} has carefully investigated the confidence interval construction for $[\beta^{\one}]^{\intercal}\Sigma\beta^{\two}$ in consideration of several high-dimensional logistic regression models. Moreover, we might need to estimate and make inferences for $[\beta^{\one}]^{\intercal}A\beta^{\two}$ with $A$ denoting a general matrix. The idea in \eqref{eq: debiased inner} can be generalized to make inference for $[\beta^{\one}]^{\intercal}A\beta^{\two}.$ The work \cite{guo2020inference} applied such a generalized debiased estimator of $[\beta^{\one}]^{\intercal}A\beta^{\two}$ in an intermediate step to determine the optimal aggregation weight of multiple regression models. 
	
	%%%%%%%%%%%%%%%%%%%%%%%%%%%%
	\subsection{R package \texttt{SIHR}}
	\label{sec: package}
	%%%%%%%%%%%%%%%%%%%%%%%%%%%%
	%\Tony{Zijian, it might be better to put the discussion about implementation and the R package \texttt{SIHR} in one single place, instead of scattering around.}\Zijian{Tony, I have collected all related package materials here. Please double check.}

	The methods reviewed in Sections \ref{sec: general idea} and \ref{sec: functional} have been implemented in the R package \texttt{SIHR} \cite{rakshit2021sihr}, which is available from CRAN. The \texttt{SIHR} package contains the main functions: \texttt{LF}, \texttt{QF}, and \texttt{CATE}. The \texttt{LF} function implements the confidence interval for $\xnew^{\intercal}\beta$ in \eqref{eq: CI linear LF} under the high-dimensional linear regression with specifying \texttt{model="linear"} and the confidence intervals for $h(\xnew^{\intercal}\beta)$ in \eqref{eq: CI logistic LF} under the high-dimensional logistic regression with specifying \texttt{model="logistic"} or \texttt{model="logisticalter"}, corresponding to the linearization weighting and link-specific weighting introduced in Section \ref{sec: debias logistic}, respectively.
	
	The \texttt{QF} function implements the confidence interval construction in \eqref{eq: CI QF linear} and hypothesis testing in \eqref{eq: test QF linear} with different choices of the weighting matrix $A$. The \texttt{CATE} function implements the confidence interval in \eqref{eq: CI linear CATE} and hypothesis testing in \eqref{eq: test logistic}. Both  \texttt{QF} and \texttt{CATE} functions can be applied to the logistic regression setting by specifying \texttt{model="logistic"} or \texttt{model="logisticalter"}. The detailed usage of the \texttt{SIHR} package can be found in the paper \cite{rakshit2021sihr}.
	
	%\Zijian{Stop here.}
	
	%The R package \texttt{SIHR} \cite{rakshit2021sihr} implements the confidence interval in \eqref{eq: CI logistic coef} with both linearization and link-specific weighting.The link-specific weight $W_k=1$ proposed in \cite{cai2021statistical} can also be adopted in constructing the point estimator in \eqref{eq: bias correction general} and the confidence interval in \eqref{eq: CI logistic LF}. The linearization weighting and link-specific weighting have been implemented under the \texttt{LF} function under the package \texttt{SIHR} with specifying \texttt{model="logistic"} or \texttt{model="logisticalter"}, respectively. 

	%%%%%%%%%%%%%%%%%%%%%%%%%%%%
	%%%%%%%%%%%%%%%%%%%%%%%%%%%%
	%\subsection{Accuracy Assessment}
	%%%%%%%%%%%%%%%%%%%%%%%%%%%%
	%%%%%%%%%%%%%%%%%%%%%%%%%%%%
	
	%%%%%%%%%%%%%%%%%%%%%%%%%%%%%%%%%%%
	%\section{Demonstration with \texttt{SIHR}}
	%%%%%%%%%%%%%%%%%%%%%%%%%%%%%%%%%%%
	
%%%%%%%%%%%%%%%%%%%%%%%%%%%%%%%%%%
%%%%%%%%%%%%%%%%%%%%%%%%%%%%%%%%%%
\section{Multiple Testing} %for High-dimensional Regression Models}
\label{sec.multiple}
%%%%%%%%%%%%%%%%%%%%%%%%%%%%%%%%%%
%%%%%%%%%%%%%%%%%%%%%%%%%%%%%%%%%%

In the previous sections, we have examined statistical inference for individual regression coefficients and related one-dimensional functionals. However, in many applications, such as genomics, it is necessary to perform simultaneous inference for multiple regression coefficients while controlling for the false discovery rate (FDR) and false discovery proportion (FDP). In this section, we will explore several large-scale multiple testing procedures for high-dimensional regression models. We start with the linear models in Section \ref{sec.test.prob} and extend the discussion to logistic models in Section \ref{multiple.logistic.sec}. The testing procedures are unified in Section \ref{multiple.proc.sec} and the power enhancement methods are discussed next.

%%%%%%%%%%%%%%%%%%%%%%%%%%%%%%%%%%
\subsection{Simultaneous inference for linear regression}\label{sec.test.prob}
%%%%%%%%%%%%%%%%%%%%%%%%%%%%%%%%%%

One-sample simultaneous inference for high-dimensional linear regression coefficients is closely related to the problem of variable selection. Common approaches for variable selection include regularization methods, such as Lasso \cite{tibshirani1996regression}, SCAD  \cite{fan2001variable}, Adaptive Lasso \cite{zou2006adaptive} and MCP \cite{zhang2010nearly}, which simultaneously estimate parameters and select features, and stepwise feature selection techniques like LARS \cite{efron2004least} and FoBa \cite{zhang2011adaptive}, which prioritize variable selection. See the discussions in \cite{liuluo2014} and references therein.  However, both of these approaches aim to find the model that is closest to the truth, which may not be achievable in practice.  Alternatively, \cite{liuluo2014} approached the problem from a multiple testing perspective and focused on controlling false discoveries rather than achieving perfect selection results. Specifically, for the high-dimensional regression model \eqref{eq: GLM} with link function $h(z)=z$, i.e.,
%\begin{eqnarray}
%\label{model-one-sample}
$Y= X\beta+\epsilon, $
%\end{eqnarray}
where  $\beta=(\beta_{1},\dots,\beta_{p+1})^{\intercal}\in \mathbb{R}^{p+1}$, $X=(X_{1,\cdot}^{\intercal},\ldots,X_{n,\cdot}^{\intercal})^{\intercal}$, $Y=(Y_{1},\ldots,Y_{n})^{\intercal}$, and $\epsilon=(\epsilon_{1},\dots,\epsilon_{n})^\intercal$, with $\{\epsilon_{k}\}$ being independent and identically distributed (i.i.d) random variables with mean zero and variance $\sigma_{\epsilon}^2$ and independent of $X_{k,\cdot}$, $k=1,\dots,n$, 
%\Zijian{Possible to only use conditional mean is zero instead of independence? If so, the model \eqref{eq: multi-regression} is exactly the same as \eqref{eq: GLM}.} 
%\Yin{I don't quite get the question, what do you mean by ``the model \eqref{eq: multi-regression} is exactly the same as \eqref{eq: GLM}''? Also, I thought the previous sections may also need the independence between $\epsilon$ and $X$ to establish asymptotic normality?} \Zijian{The asymptotic normality of the previous sections mainly requires $\E(\epsilon_i\vert X_{i,\cdot})=0$, $\E(\epsilon^2_i\vert X_{i,\cdot})=\sigma^2$, $\E(\epsilon^{2+c}_i\vert X_{i,\cdot})\leq C$ for some $C>0$, $c>0$, and sub-gaussian $\epsilon_i$. So it is ok that we assume independence between $\epsilon_i$ and $X_{i,\cdot}.$ I only wanted to double check this.}
\cite{liuluo2014} considered the following multiple testing problem,
\begin{eqnarray}
	\label{test1}
	H_{0,i}: \beta_{i}=0\mbox{  versus  }H_{1,i}: \beta_{i}\neq0, \quad i=2,\dots,p+1,
\end{eqnarray}
with the control of FDR and FDP.

In some fields, one-sample inference may not be sufficient, particularly for detecting interactions. For example, as demonstrated in  \cite{hunter2005gene}, many complex diseases are the result of interactions between genes and the environment. Therefore, it is important to thoroughly examine the effects of the environment and its interactions with genetic predispositions on disease phenotypes. When the environmental factor is a binary variable, such as smoking status or gender, interaction detection can be approached under a two-sample high-dimensional regression framework. Specifically, interactions can be identified by comparing two high-dimensional regression models as introduced in \eqref{eq: multi-regression} with identity link function, i.e.,
$
Y^{\ssd}=X^{\ssd}\beta^{\ssd}+\epsilon^{\ssd},
$
for $d=1,2$,
and recovering the nonzero components of $\beta^{\one}_{-1} - \beta^{\two}_{-1}$, where  $\beta^{\ssd}=(\beta_{1}^{\ssd},\dots,\beta_{p+1}^{\ssd})^{\intercal}\in \mathbb{R}^{p+1}$, $X^{\ssd}=(X_{1,\cdot}^{{\scriptscriptstyle (d)}\intercal},\ldots,X_{n_d,\cdot}^{{\scriptscriptstyle (d)}\intercal})^{\intercal}$, 
$Y^{\ssd}=(Y_{1}^{\ssd},\ldots,Y_{n_d}^{\ssd})^{\intercal}$,  and $\epsilon^{\ssd}=(\epsilon_{1}^{\ssd},\dots,\epsilon_{n_d}^{\ssd})^\intercal$, with $\{\epsilon_{k}^{\ssd}\}$ being i.i.d random variables with mean zero and variance $\sigma_{\epsilon^{\td}}^2$ and independent of $X_{k,\cdot}^{\ssd}$, $k=1,\dots,n_d$.  Assume that $n_1\asymp n_2$ and let $n=\max\{n_1,n_2\}$.
Then \cite{xia2018two} investigated simultaneous testing of the hypotheses
\begin{eqnarray}
	\label{test2}
	H_{0,i}: \beta_{i}^{\one}=\beta_{i}^{\two} \; \mbox{ versus } \; H_{1,i}: \beta_{i}^{\one}\neq\beta_{i}^{\two}, \quad i=2,\dots,p+1,
\end{eqnarray}
with FDR and FDP control.

In genetic association studies, it is common to measure multiple correlated phenotypes on the same individuals. To detect associations between high-dimensional genetic variants and these phenotypes, one can individually assess the relationship between each response and each covariate, as in  \eqref{test1}, and then adjust for multiplicity in the comparisons. However, as noted by \cite{Zhou2015} and \cite{Schifano2013},  jointly analyzing these phenotypic measurements may increase the power to detect causal genetic variants. Therefore, motivated by the potential to enhance power by leveraging the similarity across multivariate responses, \cite{xia2018joint} used high-dimensional multivariate regression models to address applications in which $D$ correlated responses are measured on $n$ independent individuals:
\begin{eqnarray}
	\label{model-multivariate}
	Y_{n\times D}&=& X_{n\times (p+1)}B_{(p+1)\times D}+\Upsilon_{n\times D}, 
\end{eqnarray}
where  $Y=(Y_{\cdot,1},\ldots,Y_{\cdot,D})\in \mathbb{R}^{n\times D}$, with $Y_{\cdot,d}=(Y_{1,d},\ldots,Y_{n,d})^{\intercal}$, denotes $D$ responses with $D$ fixed, %$\mu=(\mu_{\cdot,1},\ldots,\mu_{\cdot,D})\in \mathbb{R}^{n\times D}$ with $\mu_{\cdot,d}=(\mu_{1,d},\dots,\mu_{n,d})^\T$ is the mean response matrix and the rows of $\mu$ are the same, 
and $X=(X_{1,\cdot}^{\intercal},\ldots,X_{n,\cdot}^{\intercal})^{\intercal}\in \mathbb{R}^{n\times (p+1)}$ is the covariate matrix.  In \eqref{model-multivariate},
$B=(B_{\cdot,1},\ldots,B_{\cdot,D})\in \mathbb{R}^{(p+1)\times D}$, with $B_{\cdot,d}=(B_{1,d},\dots,B_{p+1,d})^{\intercal}\in \mathbb{R}^{p+1}$, represents the regression coefficient matrix, 
where $B_{i,\cdot}$ represents the regression coefficients  of the $i^{\th}$ covariate; $\Upsilon=(\epsilon_{\cdot,1},\ldots,\epsilon_{\cdot,D})\in \mathbb{R}^{n\times D}$, where  $\epsilon_{\cdot,d}=(\epsilon_{1,d},\dots,\epsilon_{n,d})^\intercal$, and $\{\epsilon_{k,d}\}$ are i.i.d random variables with mean zero and variance $\sigma_{\epsilon}^2$ and independent of $X$. 
To examine whether the $i^{\th}$ covariate is associated with any of the $D$ responses,  \cite{xia2018joint} simultaneously tested
\begin{eqnarray}
	\label{test3}
	H_{0,i}: B_{i,\cdot}=0\; \mbox{ versus } \; H_{1,i}: B_{i,\cdot}\neq 0, \quad i=2,\dots,p+1,
\end{eqnarray}
while controlling FDR and FDP. 
Because the effect of the $i^{\th}$ variable on each of the $D$ responses may share strong similarities, namely, if $B_{i,d}\neq 0$, then the rest of the entries in this row are more likely to be nonzero, a row-wise testing method using the group-wise information is more favorable than testing the significance of the matrix $B$ column by column as in testing problem \eqref{test1}.

%%%%%%%%%%%%%%%%%%%%%%%%%%%%%%%%%%
\subsubsection{Bias corrections via inverse regression}\label{sec.inverse}
%%%%%%%%%%%%%%%%%%%%%%%%%%%%%%%%%%

In the multiple testing problems discussed in Section  \ref{sec.test.prob},  our goal is to simultaneously infer the regression coefficients while controlling for error rates. Therefore, it is crucial to begin with an asymptotically unbiased estimator for each regression component. The debiasing techniques outlined in Section  \ref{sec: debias linear}  can be used to attain nearly unbiased estimates, however, as noted in \cite{liuluo2014},  the constraints or tuning parameters utilized in debiasing can significantly affect test accuracy. Additionally, the asymptotic distribution of these debiased estimators is conditional, making it challenging to characterize the dependence structure among the test statistics, which is vital for error rate control in simultaneous inference. As an alternative, we will explore an inverse regression approach in this section, which establishes the unconditional asymptotic distribution of bias corrected statistics and allows for explicit characterization of the correlation structure.  Alternatively, the debiasing method \eqref{eq: decorrelation} in \cite{zhang2014confidence,van2014asymptotically} can be used as long as the decorrelation vectors $Z_{\cdot,j}$'s  introduced in Section  \ref{debias2.sec}  are close enough to their population counterparts. This approach will be illustrated in Section \ref{multiple.logistic.sec}.

Recall that $X^{\ssd}_{k,1}=1$. To achieve bias correction, by taking testing problem \eqref{test2} as an example,
we consider the inverse regression models obtained by regressing $X_{k,i}^{\ssd}$ on $(Y_{k}^{\ssd},  \widetilde X_{k,-i}^{\ssd})$,  where $\widetilde X_{k,-i}^{\ssd} = X_{k,-\{1,i\}}^{\ssd}$, for $i=2,\dots,p+1$:
\begin{eqnarray*}
	X_{k,i}^{\one}&=&\alpha_{i}^{\one}+(Y_{k}^{\one}, \widetilde X_{k,-i}^{\one})\gamma_{i}^{\one}+\eta_{k,i}^{\one}, \quad(k=1,\dots,n_1)\cr
	X_{k,i}^{\two}&=&\alpha_{i}^{\two}+(Y_{k}^{\two}, \widetilde X_{k,-i}^{\two})\gamma_{i}^{\two}+\eta_{k,i}^{\two}, \quad(k=1,\dots,n_2)
\end{eqnarray*}
where for $d=1,2$, $\eta_{k,i}^{\ssd}$ has mean zero and variance $\sigma_{i,d}^2$ and is uncorrelated with $(Y_{k}^{\ssd},\widetilde X_{k,-i}^{\ssd})$, and the first component of $\gamma_{i}^{\ssd}=(\gamma_{i,1}^{\ssd},\dots,\gamma_{i,p}^{\ssd})^{\intercal}$ satisfies
\begin{eqnarray}
	\label{ga_id}
	%\gamma_{i}^{\d}=-\sigma_{i,d}^2(-\beta_{i}^{\d}/\sigma_{\epsilon^{\td}}^2,\beta_{i}^{\d}\beta_{-i}^{\d}^{\T}/\sigma_{\epsilon^{\td}}^2+\Omega_{i,-i}^{\d})^{\T}, 
	\gamma_{i,1}^{\ssd}=\sigma_{i,d}^2\beta_{i}^{\ssd}/\sigma_{\epsilon^{\td}}^2, \quad i=2,\dots,p+1,
\end{eqnarray}
where $\sigma_{i,d}^2=[\{\beta_{i}^{\ssd}\}^2/\sigma_{\epsilon^{\td}}^2+\omega_{i-1,i-1}^{\ssd}]^{-1}$ with $\{\Cov(X_{k,-1}^{\ssd})\}^{-1}=\Omega_d=(\omega_{i,j}^{\ssd})$. %Here, for any vector $\mu\in \mathbb{R}^p$, $\mu_{-i}$ denotes the $(p-1)$-dimensional vector formed by removing the $i^{\th}$ entry from $\mu$.  
%For any $n\times p$ matrix $A$, $A_{i,-j}$ denotes the $i^{\th}$ row of $A$ with its $j^{\th}$ entry removed, 

Note that, $r_{i}^{\ssd}=\Cov(\epsilon_{k}^{\ssd},\eta_{k,i}^{\ssd})$ can be expressed as $-\gamma_{i,1}^{\ssd}\Cov(\epsilon_{k}^{\ssd},Y_{k}^{\ssd})=-\gamma_{i,1}^{\ssd}\sigma_{\epsilon^{\td}}^2=-\sigma_{i,d}^2\beta_{i}^{\ssd}$, 
hence we can approach the debiasing of $\beta_{i}^{\ssd}$ through the debiasing of $r_{i}^{\ssd}$, and equivalently formulate the testing problem \eqref{test2} as
\begin{eqnarray*}
	\label{test2'}
	H_{0,i}: r_{i}^{\one}/\sigma_{i,1}^2= r_{i}^{\two}/\sigma_{i,2}^2, \quad i=2,\dots,p+1.
\end{eqnarray*}
The most straightforward way to estimate $r_{i}^{\ssd}$ is to use the sample covariance between the error terms, $n_d^{-1}\sum_{k=1}^{n_d}\epsilon_{k}^{\ssd}\eta_{k,i}^{\ssd}$. However, the error terms are unknown, so we first estimate them by %\Zijian{Yin, I have changed all hat to widehat.}
%\widehat{\epsilon}_{k}^{\d}&=&Y_{k}^{\d}-\bar{Y}_{d}-(X_{k,\cdot}^{\d}-\bar{X}_d)\widehat{\beta}_d\cr
%\widehat{\eta}_{k,i}^{\d}&=&X_{k,i}^{\d}-\bar{X}_{i}^{\d}-(Y_{k}^{\d}-\bar{Y}_d,(X_{k,-i}^{\d}-\bar{X}_{\cdot,-i}^{\d}))\widehat{\gamma}_{i}^{\d},
\beas
\widehat{\epsilon}_{k}^{\ssd}=Y_{k}^{\ssd}-X_{k,\cdot}^{\ssd}\widehat{\beta}^{\ssd}, \quad
\widehat{\eta}_{k,i}^{\ssd}=X_{k,i}^{\ssd}-(Y_{k}^{\ssd}-\bar{Y}^{\ssd}, \widetilde X_{k,-i}^{\ssd})\widehat{\gamma}_{i}^{\ssd},
\eeas
where $\widehat{\beta}^{\ssd}=(\widehat{\beta}_{1}^{\ssd},\dots,\widehat{\beta}_{p+1}^{\ssd})$ and $\widehat{\gamma}_{i}^{\ssd}=(\widehat{\gamma}_{i,1}^{\ssd},\dots,\widehat{\gamma}_{i,p}^{\ssd})$ are respectively the estimators of $\beta^{\ssd}$ and $\gamma_{i}^{\ssd}$ that satisfy
\bea
\label{beta_eta1}
\max\{\vert\widehat{\beta}_{-1}^{\ssd}-\beta_{-1}^{\ssd}\vert_1,\max_{i=1,\ldots,p}\vert\widehat{\gamma}_{i}^{\ssd}-\gamma_{i}^{\ssd}\vert_1\}&=&O_{\PP}(a_{n1}),\cr
\max\{\vert\widehat{\beta}_{-1}^{\ssd}-\beta_{-1}^{\ssd}\vert_2,\max_{i=1,\ldots,p}\vert\widehat{\gamma}_{i}^{\ssd}-\gamma_{i}^{\ssd}\vert_2\}&=&O_{\PP}(a_{n2}),
\eea
for some $a_{n1}$ and $a_{n2}$ such that
\beq
\label{beta_eta2}
\max\{a_{n1}a_{n2}, a_{n2}^2\}=o\{(n\log p)^{-1/2}\}, \mbox{ and }a_{n1}=o(1/\log p).
\eeq
As noted by \citep{liuluo2014,xia2018two}, estimators $\widehat{\beta}^{\ssd}$ and $\widehat{\gamma}_{i}^{\ssd}$ that satisfy \eqref{beta_eta1} and \eqref{beta_eta2} can be obtained easily via standard methods such as the Lasso and Danzig selector. 
Following that, a natural estimator of $r_{i}^{\ssd}$ can be constructed by
$
\widetilde{r}_{i}^{\ssd}=n_d^{-1}\sum_{k=1}^{n_d}\widehat{\epsilon}_{k}^{\ssd}\widehat{\eta}_{k,i}^{\ssd}.
$
However, the bias of $\widetilde{r}_{i}^{\ssd}$ exceeds the desired rate $(n_d\log p)^{-1/2}$ for the subsequent analysis. Hence, the difference of $\widetilde{r}_{i}^{\ssd}$ and $n_d^{-1}\sum_{k=1}^{n_d}\epsilon_{k}^{\ssd}\eta_{k,i}^{\ssd}$ is calculated, and it is equal to $\widehat{\sigma}_{\epsilon^{\td}}^2\widehat{\gamma}_{i,1}^{\ssd}+\widehat{\sigma}_{i,d}^2\widehat{\beta}_{i}^{\ssd}$ up to order $(n_d\log p)^{-1/2}$ under regularity conditions, where $\widehat{\sigma}_{\epsilon^{\td}}^2=n_d^{-1}\sum_{k=1}^{n_d}\{\widehat{\epsilon}_{k}^{\ssd}\}^2$ and $\widehat{\sigma}_{i,d}^2=n_d^{-1}\sum_{k=1}^{n_d}\{\widehat{\eta}_{k,i}^{\ssd}\}^2$ are the sample variances. Hence, a bias-corrected estimator for $r_{i}^{\ssd}$ is defined as
\beq 
\label{correction}
\widehat{r}_{i}^{\ssd}=\widetilde{r}_{i}^{\ssd}+\widehat{\sigma}_{\epsilon^{\td}}^2\widehat{\gamma}_{i,1}^{\ssd}+\widehat{\sigma}_{i,d}^2\widehat{\beta}_{i}^{\ssd}.
\eeq
For the other two testing problems, the bias corrections can be performed almost exactly the same via the inverse regression technique above that translates the debiasing of regression coefficients to the debiasing of residual covariances. Note that, through such an inverse regression approach, one can appropriately deal with the dependency of the component-wise debiased statistics, which is important for the following adjustment of multiplicity and the goal of FDR control.

%%%%%%%%%%%%%%%%%%%%%%%%%%%%%%%%%%
\subsubsection{Construction of test statistics}\label{multiple.stat.sec}
%%%%%%%%%%%%%%%%%%%%%%%%%%%%%%%%%%

We next construct test statistics for each of the three problems discussed in Section \ref{sec.test.prob}, using the bias-corrected statistics as a starting point.

For problem \eqref{test1}, because testing whether $\beta_i=0$ is equivalent as testing whether the residual covariance is equal to zero, the test statistics can be constructed directly based on the bias corrected $\widehat r_i$ (it can be obtained exactly the same as $\widehat r_{i}^{\ssd}$ as shown in Section \ref{sec.inverse} where the superscript $(d)$ is dropped since there is only one sample). Then the test statistics that standardize $\widehat r_i$'s are obtained by
\beq
\label{Wi1}
W_i=\frac{\widehat r_i}{(\widehat\sigma^2_\epsilon \widehat\sigma^2_{i}/n)^{1/2}}, \quad i=2,\dots,p+1,
\eeq
where $\widehat\sigma^2_\epsilon$ and $\widehat\sigma^2_{i}$ are again the sample variances by respectively dropping the superscript $(d)$ and subscript $d$ in the one-sample case. As shown in \cite{liuluo2014}, the statistics $W_i$'s are asymptotically normal under the null. %Moreover, \cite{liuluo2014}  investigated some transformations of $W_i$'s as well.

The above construction cannot be directly applied to the problem \eqref{test2}, 
because $\beta_{i}^{\ssd}$ is not necessary equal to 0 under the two-sample null and $r_{i}^{\one}/\sigma_{i,1}^2=r_{i}^{\two}/\sigma_{i,2}^2$ is not equivalent to $r_{i}^{\one}=r_{i}^{\two}$. Thus, it is necessary to construct testing procedures based directly on estimators of $r_{i}^{\one}/\sigma_{i,1}^2-r_{i}^{\two}/\sigma_{i,2}^2$.
By the bias correction in Section \ref{sec.inverse}, \cite{xia2018two} proposed an estimator of  $ r_{i}^{\ssd}/\sigma_{i,d}^2$:
\beas
T_{i}^{\ssd}=\widehat{r}_{i}^{\ssd}/\widehat{\sigma}_{i,d}^2,\quad i=2,\dots,p+1;  d=1,2,
\eeas
and tested \eqref{test2} via the estimators $\{T_{i}^{\one}-T_{i}^{\two}: i=2,\dots,p+1\}$.
Due to the heteroscedasticity, \cite{xia2018two} considered a standardized version of $T_{i}^{\one}-T_{i}^{\two}$.  Specifically, let 
\begin{equation*}
{\small\widetilde{U}_{i}^{\ssd}=(\beta_{i}^{\ssd}+U_{i}^{\ssd})/\sigma_{i,d}^2, \mbox{ with } U_{i}^{\ssd}=n_d^{-1}\sum_{k=1}^{n_d}\{\epsilon_{k}^{\ssd}\eta_{k,i}^{\ssd}-\ep(\epsilon_{k}^{\ssd}\eta_{k,i}^{\ssd})\}.}
\end{equation*}
It was shown in \cite{xia2018two} that $T_{i}^{\ssd}$ is close to $\widetilde{U}_{i}^{\ssd}$ asymptotically under regularity conditions.
%$
%|T_{i}^{\d}-\widetilde{U}_{i}^{\d}|=O_{\PP}\{\beta_{i}^{\d}(\log p/n_d)^{1/2}\}+ o_{\PP}\{(n_d\log p)^{-1/2}\}
%$
%uniformly in $i=2,\dots,p+1$.
%\Zijian{need to define $O_{\PP}$ and $o_{\PP}$?}
%\Zijian{We have only defined ${U}_{i}^{\d}$ but not $\widetilde{U}_{i}^{\d}$?}
Because
$
\theta_{i}^{\ssd}= \Var(\widetilde{U}_{i}^{\ssd}) =\Var(\epsilon_{k}^{\ssd}\eta_{k,i}^{\ssd}/\sigma_{i,d}^2)/n_d=(\sigma_{\epsilon^{\td}}^2/\sigma_{i,d}^2+\{\beta_{i}^{\ssd}\}^2)/n_d,
$
it can be estimated by 
$
\widehat{\theta}_{i}^{\ssd}=(\widehat{\sigma}_{\epsilon^{\td}}^2/\widehat{\sigma}_{i,d}^2+\{\widehat{\beta}_{i}^{\ssd}\}^2)/n_d,
$
and the standardized statistics are defined by
\beq
\label{Wi2}
W_i=\frac{T_{i}^{\one}-T_{i}^{\two}}{(\widehat{\theta}_{i}^{\one}+\widehat{\theta}_{i}^{\two})^{1/2}}, \quad i=2,\dots,p+1,
\eeq
which are asymptotically normal under the null as studied in \cite{xia2018two}. 

For the multivariate testing problem \eqref{test3}, by taking advantage of the similar effect of the $i^{\th}$ variable on each of the responses,  a group lasso penalty \citep{yuan2006model} can be imposed to obtain an estimator of $B$, such that 
\beas
{\small \max_{1\leq d\leq D}\vert\widehat{B}_{-1,d}-B_{-1,d}\vert_1=O_{\PP}(a_{n1}),
\max_{1\leq d\leq D}\vert\widehat{B}_{-1,d}-B_{-1,d}\vert_2=O_{\PP}(a_{n2}),}
\eeas
for some $a_{n1}$ and $a_{n2}$ satisfying \eqref{beta_eta2}.
By \cite{lounici2011oracle}, the above rates can be fulfilled if the row sparsity of $B$ satisfies $s(p)=o(n^{1/3}/{\log p})$. Then following the same bias correction strategy as described in Section \ref{sec.inverse}, a  debiased estimator for $r_{i}^{\ssd}$ can be obtained via
$
\widehat{r}_{i}^{\ssd}=\widetilde{r}_{i}^{\ssd}+\widehat{\sigma}_{\epsilon}^2\widehat{\gamma}_{i,1}^{\ssd}+\widehat{\sigma}_{i,d}^2\widehat{B}_{i,d}.
$
Then similarly as the aforementioned two problems, the standardized statistic can be constructed by
\beas
\widetilde T_{i}^{\ssd}={T_{i}^{\ssd}}/{\{\widehat{\theta}_{i}^{\ssd}\}^{1/2}}, \quad i=2,\dots,p+1; d=1,\ldots,D,
\eeas
where $T_{i}^{\ssd}=\widehat{r}_{i}^{\ssd}/\widehat{\sigma}_{i,d}^2$ and $\widehat{\theta}_{i}^{\ssd}=(\widehat{\sigma}_{\epsilon}^2/\widehat{\sigma}_{i,d}^2+\widehat{B}_{i,d}^2)/n.$
Finally, a sum-of-squares-type test statistic for testing the $i^{\th}$ row of $B$ is proposed:
\beq
\label{Wi3}
W_i=\sum_{d=1}^D\{\widetilde T_{i}^{\ssd}\}^2,\quad i=2,\dots,p+1,
\eeq
which is asymptotically $\chi^2_D$ distributed under the null as studied in \cite{xia2018joint}.

%%%%%%%%%%%%%%%%%%%%%%%%%%%%%%%%%%
\subsection{Simultaneous inference for logistic regression}\label{multiple.logistic.sec}
%%%%%%%%%%%%%%%%%%%%%%%%%%%%%%%%%%
The principles of simultaneous inference for high-dimensional linear regression can also be applied to high-dimensional logistic regression models.
In particular,  \cite{ma2021global} considered
the regression model \eqref{eq: GLM}, i.e.,
$Y_k = h(X_{k,\cdot}^{\intercal}\beta)+\epsilon_k$, $k=1,\ldots,n$, with the link function $h(z)=\exp(z)/[1+\exp(z)]$, and studied the simultaneous testing problem \eqref{test1} as described in Section \ref{sec.test.prob}, namely testing
\beas
	%\label{test4}
	H_{0,i}: \beta_{i}=0\mbox{  versus  }H_{1,i}: \beta_{i}\neq0, \quad i=2,\dots,p+1,
\eeas
with FDR control.

Based on the regularized estimator $\widehat\beta$ given in \eqref{eq: penalized MLE},  \cite{ma2021global} corrected the bias of $\widehat\beta$ via the Taylor expansion of $h(u_k)$ at $\widehat u_k$ for $u_k = X_{k,\cdot}^{\intercal}\beta$ and $\widehat u_k = X_{k,\cdot}^{\intercal}\widehat\beta$, and obtained that
\beq\label{approx.linear}
Y_k - h(\widehat u_k) + h'(\widehat u_k)X_{k,\cdot}^{\intercal}\widehat\beta = h'(\widehat u_k)X_{k,\cdot}^{\intercal}\beta + (R_k + \epsilon_k),
\eeq
where $R_k$ is the remainder term as specified in \eqref{eq: taylor}. Next, $Y_k - h(\widehat u_k) + h'(\widehat u_k)X_{k,\cdot}^{\intercal}\widehat\beta$ can be treated as the new response, $h'(\widehat u_k)X_{k,\cdot}$ as the new covariates, and $R_k + \epsilon_k$ as the new noise. Under such a formulation, testing $H_{0,i}: \beta_{i}=0$ can be translated into the simultaneous inference for the regression coefficients of an approximate linear model. \cite{ma2021global} applied the decorrelation method \eqref{eq: decorrelation} and constructed the following debiased estimator $\widetilde \beta_i^{\rm Deb}$,
\beq\label{Deb.logistic}
\widetilde \beta_i^{\rm Deb} = \widehat\beta_i + \frac{\sum_{k=1}^n Z_{k,i}(Y_k - h(X_{k,\cdot}^{\intercal}\widehat\beta))}{\sum_{k=1}^n Z_{k,i} h'(X_{k,\cdot}^{\intercal}\widehat\beta)X_{k,i}}, \quad i=2,\dots,p+1,
\eeq 
where $Z_{\cdot,i}$ is determined by the scaled residual that regresses $X_{\cdot,i}$ on $X_{\cdot, -i}$ through the linearization weighting $\widehat W_k = 1/h'(X_{k,\cdot}^{\intercal}\widehat\beta)$ introduced in Section \ref{sec: debias logistic}; same strategy was also employed in \citep{ren2016asymptotic,cai2019differential}. 
As summarized in Section \ref{sec.inverse},  for the subsequent FDR control analysis, the decorrelation vectors $Z_{\cdot,i}$'s were shown to be close to the true regression errors. 
Alternatively, the inverse regression technique in Section \ref{sec.inverse} can be similarly applied in the approximate linear model \eqref{approx.linear} for the bias correction.

Based on \eqref{Deb.logistic}, \cite{ma2021global} proposed the following standardized test statistic:
\beq
\label{Wi4}
W_i=\frac{\widetilde \beta_i^{\rm Deb}}{\{\sum_{k=1}^nh'(\widehat u_k)Z_{k,i}^2\}^{1/2}/\{\sum_{k=1}^nh'(\widehat u_k)Z_{k,i}X_{k,i}\}},\quad i=2,\dots,p+1,
\eeq
which is asymptotically normal under the null. 
Additionally, \cite{ma2021global} extended the idea to the two-sample multiple testing \eqref{test2}. Similarly,  multiple testing of \eqref{test3} can also be approached in the logistic setting.

%%%%%%%%%%%%%%%%%%%%%%%%%%%%%%%%%%
\subsection{Multiple testing procedure}\label{multiple.proc.sec}
%%%%%%%%%%%%%%%%%%%%%%%%%%%%%%%%%%

Using the test statistics $\{W_i: i=2,\dots,p+1\}$ in \eqref{Wi1}, \eqref{Wi2}, \eqref{Wi3} and \eqref{Wi4} as a foundation, we next examine a unified multiple testing procedure that guarantees error rates control. 

Let $\cH=\{2,\dots,p+1\}$, $\mathcal{H}_0$ be the set of true null indices and $\cH_1=\cH\setminus\cH_0$ be the set of true alternatives. We are interested in cases where most of the tests are nulls, that is, $\vert\cH_1\vert$ is relatively small compared to $\vert\cH\vert$. 
Let $\Psi(\cdot)$ be the asymptotic cumulative distribution function (cdf) of $W_i$ under the null and let $\Phi(\cdot)$ be the standard normal cdf, then we develop a normal quantile transformation of $W_{i}$ by $N_i = \Phi^{-1}\left\{1-(1- \Psi(W_i))/2\right\}$,
which approximately has the same distribution as the absolute value of a standard normal random variable under the null $H_{0,i}$. Let $t$ be the threshold level such that $H_{0,i}$ is rejected if $N_i\geq t$. 
For any given $t$, denote by 
$
R_{0}(t) = \sum_{i\in \mathcal{H}_0}\bI\{N_i\geq t\}
$
and  
$
R(t)= \sum_{i\in \cH}\bI\{N_i\geq t\}
$ %\Zijian{I have changed the indicator function to $\bI$. Please double check.}
the total number of false positives and the total number of rejections, respectively.
Then the FDP and FDR are defined as 
\[
\text{FDP}(t)=\frac{R_{0}(t)}{\max\{R(t), 1\}}, \quad \text{FDR}(t)=\mathbb{E}\{\text{FDP}(t)\}.
\]
An ideal choice of $t$ rejects as many true positives as possible while controlling the FDP at the pre-specified level $\alpha$, i.e.,
\[
t_0=\inf\left\{0\leq t\leq (2\log p)^{1/2}: \; \text{FDP}(t)\leq \alpha\right\}.
\]
Since $R_{0}(t)$ can be estimated by $2\{1-\Phi(t)\}\vert\mathcal{H}_0\vert$ and $\vert\mathcal{H}_0\vert$ is upper bounded by $p$, we conservatively estimate it by $2p\{1-\Phi(t)\}$. Therefore, the following multiple testing algorithm is proposed in \citep[e.g.,][]{liuluo2014, xia2018two, xia2018joint, ma2021global}. 

\begin{algorithm}[h]
	\caption{The multiple testing procedure.}
	\label{alg1}{
		\begin{description}
			\item[\textnormal{\bf Step 1.}] Obtain the transformed statistics $N_i=\Phi^{-1}\left\{1-(1- \Psi(W_i))/2\right\}$ from the test statistics $W_i$, $i=2,\dots,p+1$.
			
			\item[\textnormal{\bf Step 2.}]For a given $0\leq \alpha\leq 1$, calculate
			\beq\label{t_hat}
			\small
			\widehat{t}=\inf\left[0\leq t\leq (2\log p-2\log\log p)^{1/2}: \;  \frac{2p\{1-\Phi(t)\}}{\max\{R(t), 1\}}\leq \alpha\right].
			\eeq
			If (\ref{t_hat}) does not exist, then set $\widehat{t}=(2\log p)^{1/2}$. 
			
			\item[\textnormal{\bf Step 3.}] For $i\in \cH$, reject $H_{0,i}$ if $N_i\geq \widehat{t}$.
	\end{description} }
\end{algorithm}
As noted in \cite{xia2018multiple}, the constraint $0\leq t\leq (2\log p-2\log\log p)^{1/2}$ in \eqref{t_hat} is critical. When $t$ exceeds the upper bound, $2p\{1-\Phi(t)\}\rightarrow 0$ is not even a consistent estimate of $R_0(t)$. However, Benjamini-Hochberg (B-H) procedure \citep{BenHoc95} used $2p\{1-\Phi(t)\}$ as an estimate of $R_0(t)$ for all $t\geq 0$ and hence may not be able to control the FDP.
% because $|{R_0(t)}/\{2p\{1-\Phi(t)\}-1|\not\rightarrow 0$  in probability as $(n,p)\rightarrow \infty$. %However, direct application of the the Benjamini-Hochberg (B-H) procedure \citep{BenHoc95} amounts to using $2p\{1-\Phi(t)\}$ as an estimate of $R_0(t)$ for all $t\geq 0$, and it may not be able to control  the FDP. 
% For example, when the number of true alternatives is fixed, it is shown  in Proposition 2.1 in \cite{liu2013phase} that the B-H procedure cannot control the FDP with positive probability.  
On the other hand, if $t$ is not attained in the range, it is important to threshold $N_i$ at $(2\log p)^{1/2}$, because thresholding $N_i$ at $(2\log p-2\log\log p)^{1/2}$ may cause too many false rejections. 
As a result, by applying the above multiple testing algorithm to each of the problems in Sections \ref{sec.test.prob} and \ref{multiple.logistic.sec}, under some regularity conditions as specified in \citep{liuluo2014, xia2018two, xia2018joint,ma2021global}, we reach both the FDP and FDR control at the pre-specified level $\alpha$ asymptotically, i.e.,  $\lim_{(n,p)\rightarrow \infty}\PP\{\text{FDP}_{\text{Alg1}}\leq \alpha+\epsilon\}=1$ for any $\epsilon>0$ and $\limsup_{(n,p)\rightarrow \infty}\text{FDR}_{\text{Alg1}}\leq \alpha$.

%%%%%%%%%%%%%%%%%%%%%%%%%%%%%%%%%%
\subsection{Power enhancement}
%%%%%%%%%%%%%%%%%%%%%%%%%%%%%%%%%%

In addition to controlling error rates, we consider strategies to improve the power of multiple testing procedures, focusing on enhancing the power for two-sample inference when the high-dimensional objects of interest are individually sparse. This is explored in \cite{xia2020gap} with an extension to a more general framework in \cite{liang2022locally}. It is worth noting that \cite{xia2020gap}  improved the performance of Algorithm 1 by utilizing unknown sparsity in two-sample multiple testing. Additionally, power enhancement for the simultaneous inference of GLMs can be achieved through data integration  \citep[e.g.,][]{cai2021individual,liu2021integrative} as well as other power boosting methods designed for general multiple testing problems as introduced in Section \ref{intro.sec}.

Recall that, in the two-sample problem studied in Section \ref{sec.test.prob}, we aim to make inference for $\delta_i=\bI \{\beta_{i}^{\one}\neq \beta_{i}^{\two}\}$, $i=2,\dots,p+1$. %, where $\bI(\cdot)$ is an indicator function. 
Following Algorithm \ref{alg1}, we first summarize the data by a single vector of test statistics $\{N_2, \cdots, N_{p+1}\}$ and then choose a significance threshold to control the multiplicity. However, such approach ignores the important feature that both objects $\beta^{\one}$ and $\beta^{\two}$ are individually sparse. Let $\mathcal I_d=\{i=2,\dots,p+1: \beta_{i}^{\ssd}\neq 0\}$ denote the support of $\beta^{\ssd}$, $d=1, 2$, and $\mathcal I=\mathcal I_1 \cup \mathcal I_2$  the union support. Because the small cardinality of $\mathcal I$ implies that both $\beta^{\one}$ and $\beta^{\two}$ are sparse, the information on $\mathcal I$ can be potentially utilized to narrow down the alternatives via the logical relationship that
$
\mbox{$i\notin \mathcal I$ implies that $\delta_i=0$. }
$

The goal of \cite{xia2020gap} is to incorporate the sparsity information to improve the testing efficiency, and it is accomplished via the construction of an additional covariate sequence $\{S_i: i=2,\dots,p+1\}$ to capture the information on $\mathcal I$.
Note that $S_i$ and $N_i$ have different roles: $N_i$ is the primary statistic to evaluate the significance of the test, while $S_i$ is the auxiliary one that captures the sparsity information to assist the inference. It is also important to note that $S_i$ should be asymptotically independent with $N_i$ so that the null distribution of $N_i$ would not be distorted by the incorporation of $S_i$. For the two-sample problem in Section \ref{sec.test.prob}, such auxiliary statistics can be constructed by
\beq\label{S-Reg}
\small
S_i=\frac{{\widehat r_{i}^{\one}}/{\widehat{\sigma}_{{i,1}}^2} + (\widehat{\theta}_{i}^{\one}/\widehat{\theta}_{i}^{\two})({\widehat r_{i}^{\two}}/{\widehat{\sigma}_{{i,2}}^2}) }{\{{\widehat{\theta}_{i}^{\one} (1+\widehat{\theta}_{i}^{\one}/\widehat{\theta}_{i}^{\two})}\}^{1/2}}, \quad i=2,\dots,p+1. 
\eeq 
Then based on the pairs of statistics $\{(N_i,S_i): i=2,\dots,p+1\}$, the proposal in \cite{xia2020gap} operates in three steps: grouping, adjusting and pooling (GAP).  The first step divides all tests into $K$ groups based on $S_i$, which leads to heterogeneous groups with varied sparsity levels. The second step adjusts the $p$-values 
to incorporate the sparsity information.  
The final step combines the adjusted $p$-values and chooses a threshold to control the global FDR. Based on the $p$-values obtained by the test statistics $N_i$'s, i.e., $p_i = 2\{1-\Phi(N_i)\}$, the algorithm is summarized in Algorithm \ref{alg2} and we refer to \cite{xia2020gap} for its detailed implementations such as the choices of the number of groups and the grid sets.

\begin{algorithm}[t]
	\caption{Multiple testing via grouping, adjusting and pooling (GAP).}
	\label{alg2}{
		\begin{description}
			\item[\textnormal{\bf Step 1 (Grouping).}] Divide hypotheses into $K$ groups: $\mathcal G_l=\{i=2,\dots,p+1: \lambda_{l-1}<S_i\leq \lambda_l\}$, for $1\leq l\leq K$. The optimal choice of grouping will be determined in Step 2.

			\item [\textnormal{\bf Step 2 (Adjusting).}]  Define $m_l=\vert\mathcal G_l\vert$. Calculate adjusted $p$-values $p_i^w=\min\{p_i/w_l^o, 1\}$ if $i\in 
			\mathcal G_l$, $1\leq l\leq K$, where $p_i = 2\{1-\Phi(N_i)\}$ and $w_l^o$ will be calculated as follows. 
			
			\begin{itemize}
				\item \emph{Initial adjusting.} For a given grouping $\{\mathcal G_l: 1\leq l\leq K\}$, let $\widehat \pi_l$ be the estimated proportion of non-nulls in $\mathcal G_l$. Compute the group-wise weights 
				\beq\label{weight}
				\small
				w_l=\left\{\sum_{l=1}^K \frac{ m_l\widehat \pi_l}{1-\widehat \pi_l}\right\}^{-1} \frac{p \widehat \pi_l}{(1-\widehat \pi_l)}, \; 1\leq l\leq K. 
				\eeq
				Define $p_i^w=\min\{p_i/w_l,1\}$ for $i\in \mathcal G_l$.
				
				\item \emph{Further refining.} For each $\Lambda=\{\lambda_l: 1\leq l\leq K-1\}$ (allowed to be empty), let
				\beq \label{BH.proc}
				\small
				k=\max\{j: p^w_{(j)}\leq {j}\alpha/p\},
				\eeq
				and reject the hypotheses corresponding to $\{p^w_{(1)}, \ldots, p^w_{(k)}\}$, where $p^w_{(1)}\leq \cdots \leq p^w_{(p)}$ are the re-ordered $p$-values. The weights $w_l^o$ are computed using \eqref{weight} based on the {optimal} grouping that yields most rejections.  
			\end{itemize}
			%This step up-weights the hypotheses from groups with higher proportions of signals, and down-weight hypotheses from groups with lower proportions.
			
			\item [\textnormal{\bf Step 3 (Pooling).}] Combine $p_i^w$'s computed from Step 2 based on the optimal grouping, apply \eqref{BH.proc} again and output the rejections.
			
	\end{description} }
\end{algorithm}

%\Zijian{Not sure whether we need the following implementation details? We can definitely keep it if it is very important.}
%Note that, in Step 1 of Algorithm \ref{alg2},  $\lambda_0=-\infty$, $-4\sqrt{\log p}\leq\lambda_1<\lambda_2<\cdots<\lambda_{K-1}\leq4\sqrt{\log p}$, $\lambda_K=\infty$, and ${\Lambda}=\{\lambda_l: 1\leq l\leq K-1\}$ is a subset of points from a regular grid $\mathcal X=\{(j/N)\sqrt{\log p}: j=-4N,-4N+1,\ldots,-1,0,1,\ldots, 4N-1, 4N\}$, with $N$ being a large integer. There is a tradeoff in the choice of $K$ and \cite{xia2020gap}  recommends $K=3$ or 4 to balance the performance of the method and the computation loads.  
%In addition, Step 2 requires the estimation of the non-null proportion for each group, and the estimator $\widehat \pi_l^*$ proposed by \cite{schweder1982plots} and \cite{storey2002direct} is employed in \cite{xia2020gap}. 
%The final estimator sets  $\widehat\pi_l=(\epsilon\vee \widehat \pi_l^*)\wedge(1-\epsilon)$ with $\epsilon=10^{-5}$ to increase the stability of the algorithm.

In addition, \cite{xia2020gap} provided some insights on why the GAP algorithm works. First, it adaptively chooses the group-wise FDR levels via adjusted $p$-values and effectively incorporates group-wise information.
Intuitively, Algorithm \ref{alg2} increases the overall power by assigning higher FDR levels to groups where signals are more common. It does not assume known groups and searches for the optimal grouping to maximize the power. Moreover, the construction of the weights in Algorithm \ref{alg2} ensures that after all groups are combined, the weights are always ``proper'' in the sense of \cite{genovese2006false}. As a result, even inaccurate estimates of the non-null proportions would not affect the validity of overall FDR control.

Then, same as Algorithm \ref{alg1}, \cite{xia2020gap} provided the asymptotic error rates control results for Algorithm \ref{alg2}, namely, we have $\lim_{(n,p)\rightarrow \infty}\PP\{\text{FDP}_{\text{Alg2}}\leq \alpha+\epsilon\}=1$ for any $\epsilon>0$ and $\limsup_{(n,p)\rightarrow \infty}\text{FDR}_{\text{Alg2}}\leq \alpha$. Moreover, due to the informative weights \eqref{weight} that effectively incorporate the sparsity information through $\{S_i: i=2,\dots,p+1\}$, Algorithm \ref{alg2} dominates Algorithm \ref{alg1} in power asymptotically. Specifically, \cite{xia2020gap} provided the rigorous theoretical comparison that  $\Psi_{\rm Alg2}\geq \Psi_{\rm Alg1}+o(1)$ as $p\rightarrow \infty$, where $\Psi_{\rm Alg1}$ and $\Psi_{\rm Alg2}$ represent the expectations of the proportions of correct rejections among all alternative hypotheses for Algorithms \ref{alg1} and \ref{alg2}, respectively.

%%%%%%%%%%%%%%%%%%%%%%%%%%%%
%%%%%%%%%%%%%%%%%%%%%%%%%%%%
\section{Discussion}
\label{sec: discussion}
%%%%%%%%%%%%%%%%%%%%%%%%%%%%
%%%%%%%%%%%%%%%%%%%%%%%%%%%%
In this expository paper, we provided a review of methods and theoretical results on statistical inference and multiple testing for high-dimensional regression models, including linear and logistic regression. Due to limited space, we were unable to discuss a number of related inference problems. In this section, we briefly mention a few of them.

\noindent
{\bf Accuracy assessment.}
Accuracy assessment is a crucial part of high-dimensional uncertainty quantification. Its goal is to evaluate the estimation accuracy of an estimator $\widehat{\beta}$. For linear regression, \cite{cai2018accuracy} considered a collection of estimators $\widehat{\beta}$ and established the minimaxity and adaptivity of the point estimation and confidence interval for $\|\widehat{\beta}-\beta\|_{q}^{q}$ with $1\leq q\leq 2.$ Suppose that $\widehat{\beta}$ is independent of the data $\{X_{k,\cdot}, Y_k\}_{1\leq k\leq n}$ (which can be achieved by sample splitting, for example). Define the residue $Re_k=Y_k-X_{k,\cdot}^{\intercal}\widehat{\beta}=X_{k,\cdot}^{\intercal}(\beta-\widehat{\beta})+\epsilon_k$. The quadratic functional inference approach introduced in \cite{cai2020semisupervised} can be applied to the data $\{Re_k,X_{k,\cdot}\}_{1\leq k\leq n}$ to make inference for both $(\widehat{\beta}-\beta)^{\intercal}\Sigma(\widehat{\beta}-\beta)$ and $\|\widehat{\beta}-\beta\|_2^2$; see more details in Section 5.2 of \cite{cai2020semisupervised}. \cite{nickl2013confidence} used an estimator of $\|\widehat{\beta}-\beta\|_2^2$ to construct a confidence set for $\beta$ with $\widehat{\beta}$ denoting an accurate high-dimensional estimator. In the context of approximate message passing, \cite{donoho2011noise,bayati2011lasso} considered a different framework with $n/(p+1)\rightarrow \delta\in (0,1)$ and independent Gaussian design and established the asymptotic limit of $\|\widehat{\beta}-\beta\|_2^2/(p+1)$  for the Lasso estimator $\widehat{\beta}.$

\noindent
{\bf Semi-supervised Inference.} The semi-supervised inference is well motivated by a wide range of modern applications, such as Electronic Health Record data analysis. In addition to the labeled data, additional covariate observations exist in the semi-supervised setting. It is critical to leverage the information in the unlabelled data and improve the inference efficiency. As reviewed in Section \ref{sec: semi-super}, the additional unlabelled data improves the accuracy of various high-dimensional inference procedures by facilitating the estimation of $\Sigma$ or $\Sigma^{-1}$ \cite{javanmard2018debiasing,cai2020semisupervised}.
Moreover, in a different context where the linear outcome model might be misspecified, one active research direction in semi-supervised learning is to construct a more complicated imputation model (e.g., by applying the classical non-parametric or machine learning methods) and conduct a following-up bias correction after outcome imputation \citep[e.g.,][]{zhang2019semi,chakrabortty2018efficient,zhang2021double,deng2020optimal,hou2021surrogate}.

\noindent
{\bf Applications of quadratic form inference.} 
The statistical inference methods for quadratic functionals in Section \ref{sec: inner} and inner products in Section \ref{sec: QF} are useful in a wide range of statistical applications. Firstly, the group significance test of $\|\beta_{G}\|_2^2=0$ or $\beta_{G}^{\intercal}\Sigma_{G,G}\beta_{G}=0$ forms an important basis for designing computationally efficient hierarchical testing approaches \cite{mandozzi2016hierarchical,guo2021group}. Secondly, consider the high-dimensional interaction model $Y_k=A_k\eta+X_{k,\cdot}^{\intercal}\tau+A_k\cdot X_{k,\cdot}^{\intercal}\gamma$ with $A_k$ denoting the variable of interest (e.g., the treatment) and $X_{k,\cdot}$ denoting a large number of other variables. In this model, testing the existence of the interaction term can be reduced to the inference for $\|\gamma\|_2^2.$ Lastly, in the two-sample model \eqref{eq: multi-regression}, the methods proposed in Sections \ref{sec: inner} and \ref{sec: QF} have been applied in  \cite{guo2023robust} to calculating the difference $\|\beta^{\one}-\beta^{\two}\|_2^2$, which has the expression of $\|\beta^{\one}\|_2^2+\|\beta^{\two}\|_2^2-2[\beta^{\one}]^{\intercal}\beta^{\two}.$

%High-dimensional inference with unmeasured confounders is an active research area at the interaction of high-dimensional inference and causal inference. Recently, significant progress has been made in high-dimensional inference with instrumental variables; see, for example, \cite{gautier2011high,lin2015regularization,belloni2012sparse,fan2014endogeneity,guo2018confidence}. Moreover, the dense confounding structure has been utilized for deconfounding and making reliable causal inferences \cite{cevid2020spectral,guo2020doubly,bing2022adaptive}. 

\noindent
{\bf Multiple heterogeneous regression models.}
It is essential to perform efficient integrative inference in various applications that combine multiple regression models. Examples include transfer learning, distributed learning, federated learning, and distributionally robust learning. Transfer learning provides a powerful tool for incorporating data from related studies to improve estimation and inference accuracy in a target study of direct interest. \cite{li2021transfer} studied transfer learning for high-dimensional linear regression. Minimax optimal convergence rates were established, and data-driven adaptive algorithms were proposed. \cite{Li2021Transfer-GLM, tian2022transfer} explored transfer learning in high-dimensional GLMs and
\cite{battey2018distributed,lee2017communication} considered distributed learning for high-dimensional regression. 
Additionally, \cite{liu2021integrative,cai2021individual} proposed integrative estimation and multiple testing procedures of cross-sites high-dimensional regression models that simultaneously accommodated between study heterogeneity and protected individual-level data.
In a separate direction, \cite{meinshausen2015maximin} proposed the maximin effect as a robust prediction model for the target distribution being generated as a mixture of multiple source populations. \cite{guo2020inference} further established that the maximin effect is a group distributionally robust model and studied the statistical inference for the maximin effect in high dimensions. Many questions in multiple heterogeneous regression models are open and warrant future research.  

\noindent
{\bf Other simultaneous inference problems.}
The principles of multiple testing procedures reviewed in Section \ref{sec.multiple} are also widely applicable to a range of simultaneous inference problems, including graph learning \citep[e.g.,][]{liu2013ggm,xia2017hypothesis,xia2018multiple,cai2019differential}, differential network recovery \citep[e.g.,][]{xia2015testing,xia2019matrix,kim2021two}, as well as various structured regression analysis \citep[e.g.,][]{zhang2020estimation,li2021inference,sun2022decorrelating}. 
For example, through similar regression-based bias-correction techniques as reviewed in Section \ref{sec.inverse}, \cite{liu2013ggm} studied the estimation of Gaussian Graphical Models with FDR control;
\cite{xia2018multiple} focused on the recovery of sub-networks in Gaussian graphs; 
\cite{cai2019differential} identified significant communities for compositional data. Additionally,
\cite{xia2015testing,xia2019matrix} proposed multiple testing procedures for differential network detections of vector-valued and matrix-valued observations, respectively.
Multiple testing of structured regression models such as mixed-effects models and confounded linear models has also been extensively studied in the literature \citep{li2021inference,sun2022decorrelating}.

\noindent
{\bf Alternative multiple testing methods for regression models.}
Besides the bias-correction based multiple testing approaches reviewed in this article, there are a few alternative classes of methods that aim at finite sample FDR control for linear models. Examples include knockoff based methods, mirror statistics approaches and $e$-value proposals. 
In particular, \cite{barber2015controlling} constructed a set of knockoff variables and selected those predictors that have considerably higher importance scores than the knockoff counterparts; 
\cite{candes2018panning} extended the work to a model-X knockoff framework that allowed unknown conditional distribution of the response. There are several generalizations along this line of research; see \cite{barber2020robust,ren2021derandomizing} and many references therein. 
Inspired by the knockoff idea, \cite{du2021false} proposed a symmetrized data aggregation approach to build mirror statistics that incorporate data dependence structure; a general framework on mirror statistics of GLMs was studied in \cite{dai2023scale} and the references therein.
The $e$-value  based proposal  \citep{vovk2021values} is another useful tool for multiple testing in general.
\cite{wang2020false} proposed an e-BH procedure that achieved FDR control under arbitrary dependence among the $e$-values, and the equivalence between the knockoffs and the e-BH was studied in \cite{ren2022derandomized}.

\setlength{\bibsep}{1.4pt plus 0.3ex} 
\bibliography{HDRef,Multiple-Testing-ref}
	
	
\end{document}